\documentstyle[11pt,epsfig]{article}
\newcommand{\be}{\begin{equation}}
\newcommand{\ee}{\end{equation}}
\newcommand{\bea}{\begin{eqnarray}}
\newcommand{\eea}{\end{eqnarray}}
\newcommand{\beet}{\begin{equation*}}
\newcommand{\eeet}{\end{equation*}}
\newcommand{\beaet}{\begin{eqnarray*}}
\newcommand{\eeaet}{\end{eqnarray*}}
\newcommand{\bfig}{\begin{figure}}
\newcommand{\efig}{\end{figure}}
\newcommand{\bc}{\begin{center}}
\newcommand{\ec}{\end{center}}

\newcommand{\szz}{\sigma_{tt}}
\newcommand{\sxx}{\sigma_{xx}}
\newcommand{\szx}{\sigma_{tx}}
\newcommand{\sxz}{\sigma_{xt}}

\newcommand{\De}{\Delta}
\newcommand{\de}{\delta}
\newcommand{\La}{\Lambda}
\newcommand{\ga}{\gamma}
\newcommand{\gah}{\hat{\gamma}}

\newcommand{\ra}{\right >}
\newcommand{\la}{\left <}

\begin{document}

\title{Models of stress fluctuations in granular media}

\author{P. Claudin, J.-P. Bouchaud\\
Service de Physique de l'Etat Condens\'e,\\
CEA, Orme des Merisiers,\\
91191 Gif-sur-Yvette, Cedex France.\\
\\
M. E. Cates and J. P. Wittmer\\
Department of Physics and Astronomy,\\ University of Edinburgh, JCMB King's
Buildings,\\ Mayfield Road, Edinburgh EH9 3JZ, UK.}

\date{\today}
\setcounter{page}{0}
\maketitle

\vspace{1cm}

\begin{abstract}
We investigate in detail two models describing how stresses propagate and fluctuate in granular media. 
The first
one is a scalar model where only the vertical component of the stress
tensor is considered. In the continuum limit, this model is equivalent to a {\it diffusion
equation} (where the r\^ole of time is played by the vertical coordinate) plus a randomly
varying convection term. We calculate the response and correlation function of this
model, and discuss several properties, in particular related to the stress
distribution function. We then turn to the tensorial model, where the basic starting 
point is a {\it wave equation} which, in the absence of disorder, leads to a ray-like
propagation of stress. In the presence of disorder, the rays acquire a diffusive width
and the angle of propagation is shifted.  A striking feature is that
the response function becomes negative, which suggests that the contact network is mechanically
unstable to very weak perturbations. The stress correlation function reveals characteristic 
features related to the ray-like propagation, which are absent in the scalar description. Our analytical calculations are confirmed and extended by a numerical analysis of the stochastic wave equation. 
\end{abstract}

\thispagestyle{empty}

\vspace{8cm}

\newpage

\section{Introduction}

Granular media are materials where stress fluctuations are large, even 
on scales much larger than the grain size. Repeatedly pouring the very
same  amount of powder in a silo results in fluctuations of the weight supported by
the bottom plate of $20 \%$ or more \cite{Brown,Clement}. This weight furthermore changes 
very abruptly when temperature changes by only a few $^o C$, which induces only very small 
changes of the size of each grain \cite{Clement, CB}. 

More quantitative experiments were recently performed by Liu et al. \cite{Liu},
Brockbank et al. \cite{Huntley} and Mueth et al. \cite{Nagel}, where the local
fluctuations of the normal stress deep inside a silo or at the base of a sandpile
were measured (see also \cite{Baxter}, and for early qualitative experiments
\cite{Dantuetc}). It was found that the stress probability distribution is rather
broad (i.e. the relative fluctuations are of order one), decaying exponentially for
large stresses. A simple `scalar' model for stress propagation was introduced and
studied in detail \cite{Liu,Copper}, which predicts a stress probability
distribution in good agreement with experimental (and numerical) data. However,
this model only considers the {\it vertical} normal component of the stress
tensor, and discards  all the other components:  in this sense the model is scalar.

A fully `tensorial' model for stress propagation in homogeneous granular media was 
proposed in \cite{BCC,WCCB,FPA} to account for the pressure `dip' which is observed
experimentally below the apex of conical sandpiles. The most striking feature of this
model is that the stress propagation equation is a {\it wave equation}, with the
vertical axis playing the r\^ole of time. Correspondingly, the stress propagates
(in two dimensions, see \cite{BCC}) along two rays which makes a certain angle with
the  vertical axis (the `light cone'). This must be
contrasted with the scalar model, where stresses travel essentially vertically, 
which predicts a central pressure `hump' (rather than a `dip'). 

It is thus a priori not obvious that the scalar model is a suitable starting point 
for the description of fluctuations. Conversely, the influence of local 
randomness within the tensorial model was not yet investigated, and is very interesting
{\it per se}. In particular, it is important to know if and how the idea of a `light
cone' survives in the presence of disorder, and how the stress fluctuations develop. 

The aim of the present paper is to calculate analytically (in two dimensions)
the average response function (Green function) and the two-point correlation
function for the tensorial model in the presence of disorder, and to compare the
results to those obtained within a scalar description. We find that the cone
survives at small disorder (although the cone angle is shifted and acquires a non
zero width, which we compute). More surprisingly, the Green function takes
{\it negative values} \footnote{That the Green function can take negative values in the presence of inhomogeneities was already noticed within the {\sc fpa} model in \cite{FPA}.}, a feature which 
we checked numerically, and which we
discuss in detail in terms of the essential ``fragility'' of the contact network. We show that 
the two-point correlation function keeps a
signature of this cone like propagation. For large disorder however, the theory
suggests that the structure of the large scale equations could change drastically,
from an {\it hyperbolic} wave equation to an {\it elliptic} equation, akin to
(but distinct from) those appearing in elasticity theory. The interpretation of the equations however
suggests that  by the time this happens, the pile is unstable to any perturbations and 
spontaneously rearranges.

The `tensorial' stress probability distribution is investigated numerically, with 
certain results which are close to those of the scalar model. We
explain this by showing that a special case of the tensorial model actually reduces to
the superposition of {\it two} independent scalar models.

This paper is constructed as follows: in section 2, we review the properties of 
the scalar model, including results which appeared in the literature in very different
contexts (scalar diffusion in turbulence, localisation). In section 3, the `random
wave equation' for the tensorial case is motivated by a microscopic model and simulations,
and studied using
perturbation theory in the strength of the disorder. We discuss how the line shape of the
response function distorts from two delta
peaks to (eventually) one broad peak as disorder increases. 
Some generalisations of the the `random
wave equation' are considered in section 4. In section 5, we present numerical
results for the stress distribution function and compare them with the predictions of
the scalar model, and also of direct simulations of sphere packings
\cite{Radjai,Eloy}. We discuss a limit where the two models can be quantitatively
compared. Finally, in section 6, a summary of the most interesting results is given,
with suggestions of new experiments and open questions.

\section{The Scalar Model}
\label{Scalar}

\subsection{The discrete version}

\vskip 0.5cm
$\bullet$ Definition.
\vskip 0.5cm
The main assumption of the scalar model is that only the vertical normal component
 of the stress tensor $w = \sigma_{zz}$ (the `weight') needs to be 
considered.
\bfig[hbt]
\bc
\epsfysize=6cm
\epsfbox{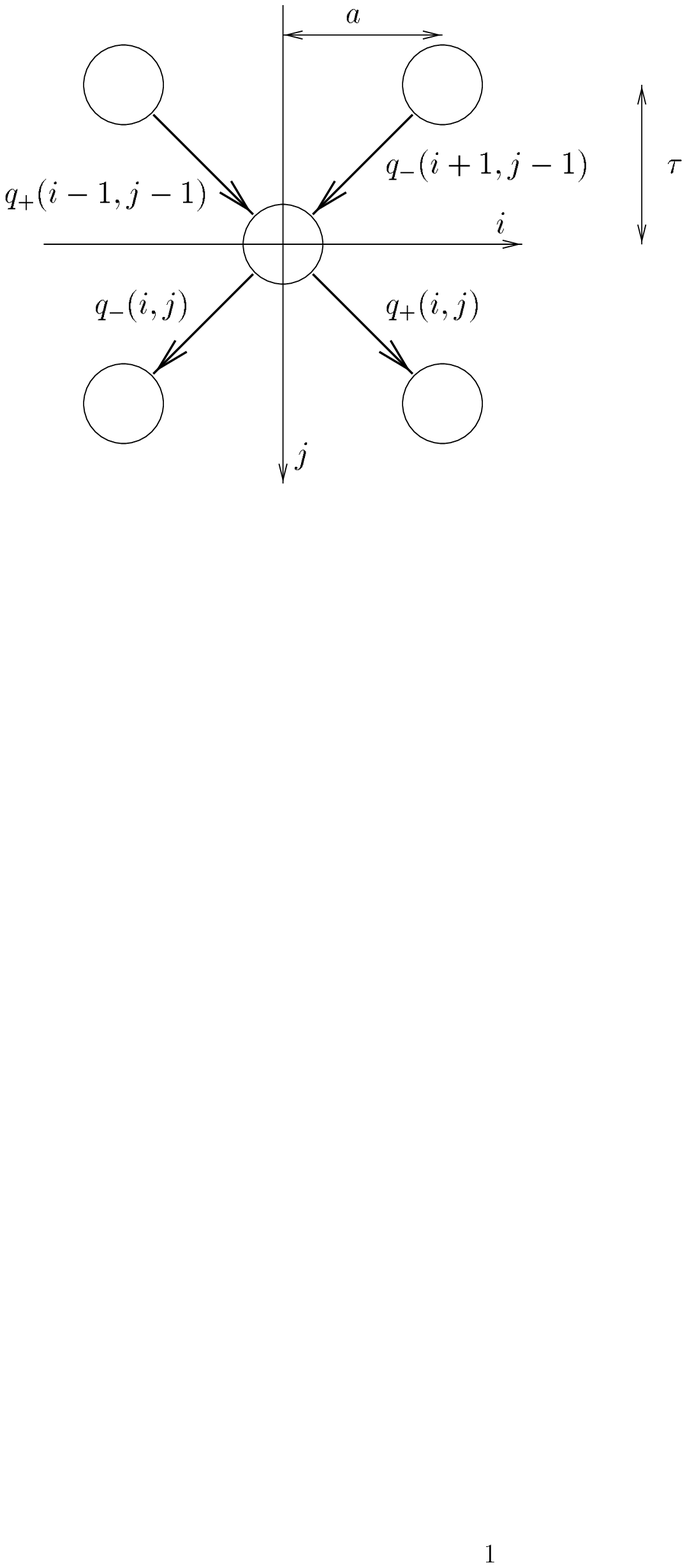}
\caption{\small The `Chicago' model with two neighbours. $q_\pm$'s are
independent random variables, except for the weight conservation constraint:
$q_+(i,j) + q_-(i,j)=1$.
\label{liufig}}
\ec
\efig
If the grains reside on the nodes of a two-dimensional lattice 
(see figure \ref{liufig}), the simplest model for weight propagation down the pile is:
\be
\label{liudiscret}
w(i,j) = w_g + q_+(i-1,j-1)w(i-1,j-1) + q_-(i+1,j-1)w(i+1,j-1)
\ee
where `$w_g$' is the weight of each grain, and $q_\pm(i,j)$ are `transmission' 
coefficients giving the fraction of weight which the grain $(i,j)$ transmits to its 
right (resp. left) neighbour immediately below.
Mass conservation imposes that $q_+(i,j) + q_-(i,j)=1$ for all $i,j$'s.
The case of an ordered pile of identical grains would correspond to 
$q_\pm = \frac{1}{2}$. The authors of \cite{Liu,Copper}
proposed to take into account (in a phenomenological way) the local disorder in
packing, grain sizes and shapes, etc., by choosing $q_+(i,j)$ to be independent
random numbers (except for the above constraint), for example uniformly distributed 
between $0$ and $1$. This model, which we shall call the 
`Chicago' model (or $q$ model), was originally written with an arbitrary number $N$
of downward neighbours ($N=2$ in the example above), and can thus be (in principle)
generalized to three dimensions.

\vskip 0.5cm
$\bullet$ Results on the stress distribution. Universality ?
\vskip 0.5cm
The case of a uniform distribution of the $q$'s is interesting because it 
leads to an exact solution for the local weight distribution $P(w)$. In this 
limit, the correlation between two neighbouring sites at the same altitude $j$ is zero
for all $j$. For more general $q$ distributions, this is true only when  $j$ is large 
(see below). Thus $P(w)$ obeys the following mean-field equation:
\be
P_{j+1}(w) = \int_0^1 dq_1 dq_2 \rho(q_1)\rho(q_2) \int_0^\infty dw_1 dw_2
P_{j}(w_1)P_{j}(w_2) \delta[w-(w_1 q_1 + w_2 q_2 +w_0)] \label{recu}
\ee
where $\rho(q)$ is the distribution of $q$, here taken to be $\rho(q) = 1$.
In the limit $j \to \infty$, the stationary distribution $P^*$ of this equation
is given by:
\be
P^*(w) = \frac{w}{W^2} \exp-\frac{w}{W}
\label{exptail}
\ee
where $2 W= j w_g$ is the average weight. For $N \neq 2$, the distribution 
is instead a $\Gamma$ distribution of parameter $N$; its small $w$ behaviour is $w^{N-1}$
while the large $w$ tail is exponential. Liu et al. \cite{Liu,Copper} have argued that
this behaviour is generic: for example, the condition for the local weight $w$ to be
small is that all the $N$ $q$'s reaching this site 
are themselves small; the phase space volume for this is proportional to $w^{N-1}$
if the distribution $\rho(q)$ is regular around $q=0$. However, if instead $\rho(q)
\propto q^{\gamma-1}$ when $q$ is small, one expects $P^*(w)$ to behave for small $w$ as $w^{-\alpha}$,
with $\alpha=1-N\gamma < 0$. Similarly, the exponential tail at large $w$ is sensitive to
the behaviour of $\rho(q)$ around $q=1$. In particular, if the maximum value
of $q$ is $q_M < 1$, one can easily show by taking the Laplace transform of equation
(\ref{recu}) that $P^*(w)$ decays {\it faster} that an exponential:
\be
\log P^*(w) \mathop{\propto}_{w \to \infty} -w^\beta
\qquad \mbox{with} \qquad
\beta=\frac{\log N}{\log (Nq_M)} 
\ee
(Notice that $\beta=1$ whenever $q_M=1$, and that $\beta \to \infty$ when $q_M=1/N$). 

In this sense, the exponential tail of $P^*(w)$ is not universal: it requires the 
possibility that one of the $q$ can be arbitrarily close to $1$. This implies that all
other $q$'s originating from that point are close to zero, i.e. that there is a nonzero
probability density that one grain is entirely bearing on one of its downward neighbours.

Note that if $q$ can only take the values $0$ or $1$, the distribution $P(w)$ becomes
a power law, $P^*(w) \propto w^{-\alpha}$, with $\alpha=4/3$ for $N=2$ \cite{Copper}.
This power law is however truncated for large $w$ as soon as values for $q$ different
from $0$ and $1$ are allowed.

How well does the simple distribution (\ref{exptail}) compare to experiments and
numerical simulations? The exponential decay for large $w$ appears in some cases to 
overestimate both the experimental \cite{Huntley} and numerical tail
\cite{Eloy} (see also section \ref{numerics}), suggesting a value of $\beta$ somewhat larger than $1$.
On the other hand, the probability to observe very small $w$ is much 
underestimated
by equation (\ref{exptail}): see \cite{Huntley,Radjai,Nagel} and section
\ref{numerics}. This might be due to the fact that {\it arching}
effects are absent in this scalar model. A generalisation of the Chicago model
allowing for arching was suggested in \cite{CB}, which generates sites where
$q_+=1$ and $q_-=0$  (or vice versa). This indeed leads to much higher probability 
density for small weights, $P^*(w) \propto w^{-\alpha}$ as argued in \cite{Copper} -- 
see also \cite{Nico}.

\subsection{Continuous limit of the scalar model.}

Let us focus on the case $N=2$ and define $v$ to be such that $q_\pm (i,j)=(1\pm v(i,j))/2$.
If $v$ is small, the local weight is smoothly varying, and the discrete equation
(\ref{liudiscret}) can then be written in the following differential form:
\be
\label{liudiff}
\partial_t w + \partial_x (vw) = \rho + D_0 \partial_{xx} w\label{eq5}
\ee
where $x=i a$ and $t=j \tau$ are the horizontal and (downwards) vertical variables
corresponding to indices $i$ and $j$ of
figure \ref{liufig}, and $a$ and $\tau$ are of the order of the size of the grains.
The vertical coordinate has been called $t$ for its
obvious analogy with time in a diffusion problem. $\rho$ is the density
of the material (the gravity $g$ is taken to be equal to $1$),
and $D_0$ a `diffusion' constant, which depends on the geometry of the
lattice on which the discrete model has been defined. For a
rectangular lattice as shown in figure \ref{liufig}, $D_0={a^2 \over 2 \tau}$.
More generally, the diffusion constant is of the order of magnitude 
of the size of the grains, $a$.

In this model and in the following, we shall assume that the density $\rho$ is not
fluctuating. Density fluctuations could be easily included; it is however easy to
understand that the resulting relative fluctuations of the weight at the bottom of
the pile decrease with the height of the pile $H$ as $H^{-1/2}$, and are thus much 
smaller than those induced by the randomly fluctuating direction of propagation,
encoded by $q$ (or $v$), which remain of order $1$ as $H \to \infty$.

Two interesting quantities to compute are the average `response' $G(x,t|x_0,t_0)$ to a
small density change at point $(x_0,t_0)$, measured at point $(x,t)$, and the correlation
function of the force field, $C(x,t,x',t') = \la w(x,t) w(x',t') \ra_c$ (connected part),
where the averaging is taken over the realisation of the noise $v(x,t)$.

Equation (\ref{liudiff}) shows that the scalar model of stress propagation is 
identical to that describing tracer diffusion in a (time dependent) flow $v(x,t)$.
This problem has been the subject of many 
recent works in the context of turbulence \cite{Russes,Verga}; we believe that interesting 
qualitative analogies with that field can be made. In particular, `intermittent' bunching of
the tracer field correspond in the present context to patches of large stresses, which may induce
anomalous scaling for higher moments of the stress field correlation function.
 We refer the reader to \cite{Russes,Verga} for further details.

\vskip 0.5cm
$\bullet$ Statistics of the noise $v(x,t)$.
\vskip 0.5cm
The noise term $v$ represents the effect of local heterogeneities in the granular packing. Its mean value
is taken to be zero, and its correlation
function is chosen for simplicity  to be of the factorable form
$\la v(x,t)v(x',t') \ra =\sigma^2 g_x(x-x') g_t(t-t')$, where $g_x$ and $g_t$ are noise
correlation functions along $x$ and $t$ axis. We shall take $g_x$ and $g_t$ to be
short-ranged (although this may not be justified: fluctuations in the microstructure of granular
media may turn out to be long-ranged due to e.g. the presence of long stress paths or
arches), with correlation lengths $\ell_x$ and $\ell_t$.
Our aim is to describe the system at a
scale $L$ much larger than both the lattice and the correlation lengths:
$a,\: \tau,\: \ell_x,\: \ell_t \ll L$. This will allow us to
look for solutions in the regime $k, E \to 0$, where $k$ and $E$ are
the conjugate variables for $x$ and $t$ respectively, in Fourier-Laplace space.
However, we shall see below that the limit $a, \tau, \ell_x, \ell_t \to 0$ can
be tricky, and must be treated with care: this is because the noise 
appears in a multiplicative manner in equation (\ref{liudiff})
\footnote{In the tensorial case, the limit $\ell_t \to 0$ actually makes the problem trivial, 
for a reason which will become clear below.}.
For computational purposes, we shall often implicitly assume that the 
probability distribution of $v$ is gaussian; this might however introduce artefacts 
which we discuss.

\vskip 0.5cm
$\bullet$ Fourier transforms.
\vskip 0.5cm

The limit where $a,\ell_x \to 0$ is ill defined and leads to a divergence
of the perturbation theory in $\sigma$ for large wavevectors $k$. 
We thus choose to regularize the problem by working within the first Brillouin
zone, i.e., we keep all wave vector components within the interval ${\cal I}=[-\La,+\La]$,
where $\La ={\pi \over a}$. Our Fourier conventions for a given quantity $f$ will
then be the following: \bea
\label{fourier_rules1}
f(x,t) & = & \int_{-\La}^{\La} {dk \over 2\pi} e^{i k x} f(k,t) \\
\label{fourier_rules2}
f(k,t) & = & \ell_x \sum_{x=-\infty}^{+\infty} e^{-i k x} f(x,t)
\eea
One has to be particularly careful when computing convolution integrals, such 
as $\int {dq \over 2\pi} f_1(q) f_2(k-q)$ which must be understood with limits $-\La+k,
\La$ (resp. $-\La, \La+k$) if $k \geq 0$ (resp. $k \leq 0$). An important example,
which will appear in the response function calculations, is:
\be
\label{convo_q}
\int_{q,k-q \in  {\cal I}}{dq \over 2\pi} q = {\Lambda k \over 2\pi} + {\cal O}(k^2)
\ee
Let us then take the Fourier transform of equation (\ref{liudiff})
along $x$, to obtain:
\be
\label{liuTF}
(\partial_t + D_0k^2) w_k = \rho_k + i k \int {dq \over 2\pi} w_q v_{k-q}
\ee
Our aim is to calculate, in the small-$k$ limit, the average response
(or Green) function $G(k,t-t')$ defined as the expectation value of the functional
derivative $\la \de w(k,t)/\de \rho (k,t') \ra$;
and the two points correlation function of $w$, $\la w(k,t)w(k',t) \ra \equiv 2\pi \de (k+k') C(k,t)$.

\vskip 0.5cm
$\bullet$ The noiseless Green function.
\vskip 0.5cm
The noiseless (bare) Green function (or `propagator') $G_0$ is the solution of the equation where
the `velocity' components $v_q$ are identically zero: 
$(\partial_t + D_0k^2) G_0(k,t-t') = \de (t-t')$
which is:
\be
\label{G01}
G_0(k,t-t') = \theta (t-t') e^{-D_0k^2(t-t')}
\ee
Or, in real space, 
\be
\label{realG0}
G_0(x,t-t') = {\theta(t-t') \over \sqrt{4\pi D_0(t-t')}}
e^{-{x^2 \over 4D_0(t-t')}}
\ee
\vskip 0.5cm
$\bullet$ Ambiguities due to  multiplicative noise. Ito vs. Stratonovitch.
\vskip 0.5cm
In equation (\ref{liuTF}), we have omitted to specify the dependence on the variable $t$.
There is actually an ambiguity in the product term $w_q v_{k-q}$. In the discrete Chicago
model \cite{Liu}, the $q_\pm$'s emitted from a given site are independent of the value of
the weight on that site. In the continuum limit, this corresponds to choosing $w_q(t)$ to
be independent of $v_{k-q}(t)$, or else that the $v$'s must be thought of as slightly
posterior to the $w$'s (i.e.  the product is read as $w_q(t-0) v_{k-q}(t+0)$).
In this case, the average of
equation (\ref{liuTF}) is trivial and coincides with the noiseless limit; hence  $G=G_0$.
This can be understood directly on the discrete model by noticing that the Green function
$G(i,j|0,0)$ can be expressed as a sum over paths, all starting at site $(0,0)$,
and ending at site $(i,j)$:
\be
\label{Q1}
G(i,j|0,0)=\sum_{\mbox{paths } \cal{P}} \ 
\prod_{\mbox{($k$,$l$)} \in \cal{P}} \: q_\pm(k,l)
\ee
where the $q_\pm(k,l)$ are either $q_+(k,l)$ or $q_-(k,l)$, depending on the  path.
Since each bond $q_\pm(k,l)$ appears only once in the product,
the averaging over $q$ is trivial and leads to:
\be
\label{Q2}
G(i,j|0,0)=\sum_{\mbox{paths } \cal{P}} \ 2^{-j} \equiv G_0(i,j|0,0)
\ee
(Note that this
argument fails for the computation of the correlation function $C$, since
paths can `interfere'. We shall return later to this calculation.)

The above choice corresponds to Ito's prescription in stochastic calculus. 
Another choice (i.e. Stratonovitch's prescription) is however possible,
which corresponds to the proper continuum time limit in the case where the 
correlation length $\ell_t$ is very small, but not smaller than $a$ (see figure \ref{gt}).
In this case, the $w$'s and the $v$'s cannot be taken to be independent.
This is the choice that we shall make in the following.
\bfig[hbt]
\bc
\epsfysize=5cm
\epsfbox{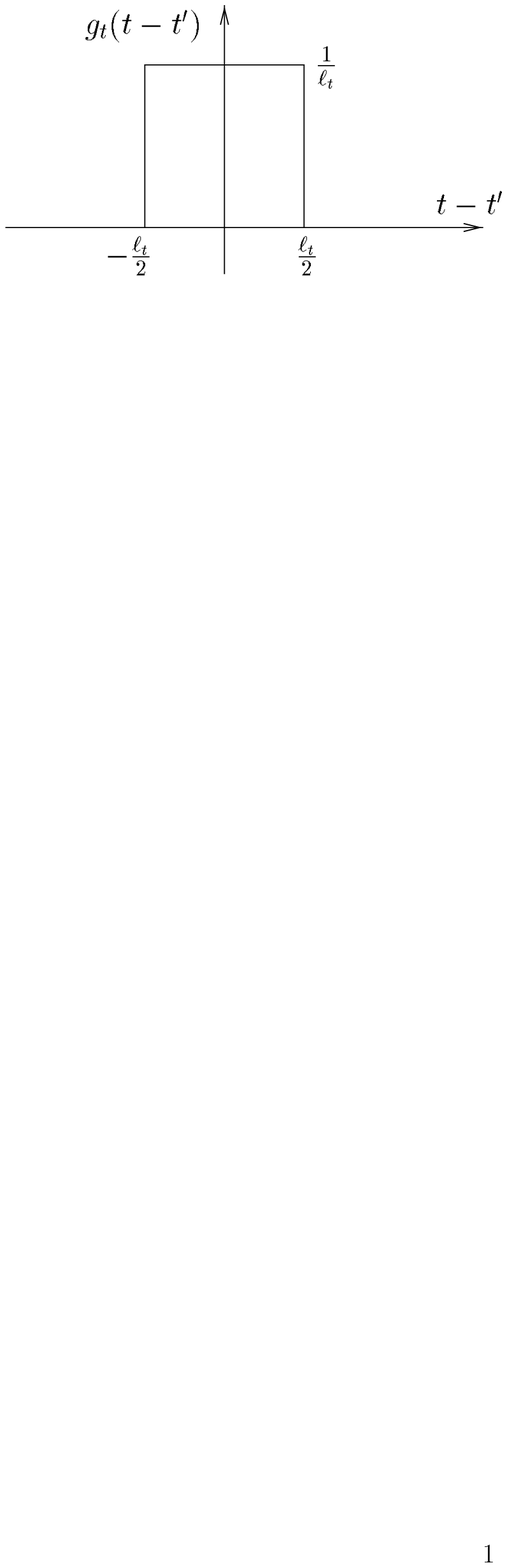}
\caption{\small Correlation function of the noise along $t$ axis. The results presented 
below would hold for an arbitrary symmetric, short range function.
\label{gt}}
\ec
\efig

\subsection{Calculation of the averaged response and correlation functions.}

Two approaches will be presented. The first one, based on
Novikov's theorem, leads to exact differential equations for $G$ and $C$,
which can be fully solved. The second one is a mode-coupling approximation ({\sc{mca}}),
based on a resummation of perturbation theory. It happens that, for this particular model
where the noise is gaussian and short range correlated in time, both approaches give
the same results, because perturbation theory is trivial. In other cases, though,
where exact solutions are no longer available, the {\sc{mca}} is in general
very useful to obtain non perturbative results (see \cite{MCA}).  

We shall see that the effect of the noise is to widen the diffusion peak:
$D_0$ is renormalized by an additional term proportional to the variance of the noise $v$.

\vskip 0.5cm
$\bullet$ Novikov's theorem. Exact equations for $G$.
\vskip 0.5cm
Novikov's theorem provides the following identity, valid if the $v$ are gaussian random variables:
\be
\label{novi1}
\la w(k,t)v(k',t) \ra =\int_0^t dt' \int dq
\left < {\de w(k,t) \over \de v(q,t')} \right > \la v(q,t')v(k',t) \ra
\ee
Such a term actually appears in equation (\ref{liuTF}), after transformation into an equation
for $G$:
\be
\label{novi2}
(\partial_t+D_0k^2)G(k,t-t')= \de (t-t')
-i k {\de \over \de \rho (k,t')} \int {dq \over 2\pi}
\la v(q,t) w(k-q,t) \ra 
\ee
In  the limit where $\ell_x = a \to 0$, the noise correlation is of the form:
$\la v(q,t)v(q',t') \ra =2 \pi \sigma^2 \de (q+q') \tilde g_x(q) g_t(t-t')$, with $g_t$
peaked in $t=t'$ such that $f(t')g_t(t-t') \simeq f(t)g_t(t-t')$ for
any function $f$. In all section \ref{Scalar} we take $g_x(q)=1$.
From formally integrating equation (\ref{liuTF}) between $t'$ and $t$,
one can express the equal-time derivative $\de w / \de v$ as:
\be
\label{novi3}
\left.{\de w(k,t) \over \de v(k',t')}\right|_{t' = t-0}= -i k w(k-k',t)
\ee
and thus obtain:
\be
\label{novi4}
(\partial_t+D_0k^2)G(k,t-t') = \de (t-t')
- 2 \pi \sigma^2 k G(k,t-t') \int_0^t dt' g_t(t-t') \int {dq \over 2\pi} (k-q)
\ee
Using the shape of the function $g_t$ (see figure \ref{gt}), the first integral is $1/2$.
The second one is a convolution integral, and its value is
$\Lambda k/2\pi + {\mathcal{O}} (k^2)$ (see equation (\ref{convo_q})). The final
differential equation for $G$ is then, in the small-$k$ limit, 
a diffusion equation with a renormalized diffusion constant:
\be
\label{Dchapeau}
D_R = D_0 + {\sigma^2 \Lambda \over 2}
\ee
It is interesting to note that the model remains well defined in the limit where the
`bare' diffusion constant is zero, since a non zero diffusion constant is induced by
the fluctuating velocity $v$. This would not be true if equation (\ref{eq5}) was interpreted with the Ito convention,
where the fluctuating velocity would {\it not} lead to any spreading of the average density.

The most important conclusion is thus that, in the present scalar model, stresses propagates
essentially vertically: taking $\ell \sim a$, the response at depth $H$ to a small perturbation
is confined within a distance $\propto \sqrt{D_R H}$ from the vertical. Since
$D_R \simeq \ell^2/a$, $\sqrt{D_R H}$ is much less than $H$ in the limit where
$H \gg \ell^2/a$, i.e. when the height of the assembly of grains is much
larger than the grain size.
 
\vskip 0.5cm
$\bullet$ Exact equations for $C$.
\vskip 0.5cm
Exact equations can also be derived for $C$, following very similar
calculations. From equation (\ref{liuTF}), one can deduce the corresponding
one for $w(k,t)w(-k,t)$. Upon averaging, Novikov's theorem has
to be used on quantities such as $\la w(k,t)v(q,t)w(-k-q,t) \ra$, finally leading to:
\be
\label{C1}
(\partial_t+2D_R k^2) C(k,t) = \sigma^2 k^2 \int {dq \over 2\pi} C(q,t)
\ee
One can formally integrate equation (\ref{C1}). It gives
\be
\label{C2}
C(k,t)= C(k,0) e^{-2D_R k^2 t}+
\sigma^2 k^2 \int_0^t dt' e^{-2D_R k^2 (t-t')} \tilde{C}(t')
\ee
where $\tilde{C}(t')=\int {dk \over 2\pi} C(k,t')$.
Let us specify at this stage two specific initial conditions $C(k,0)$ which can be of interest.
We consider, for simplicity, a random packing of `table tennis' balls with no mass ($\rho =0$),
but subject to a random overload of zero mean
($\la w(x, 0)w(x', 0) \ra = A_0^2 \de (x-x')$) or to a constant overload ($w(x,t=0)=B_0$).
Therefore, $C(k,0)=A_0^2$ in the first case, and
$C(k,0)=B_0^2 \de (k)$ in the second one. Equation (\ref{C2})
is then solved in two steps: we first integrate it over $k$ and find a closed
equation for $\tilde{C}$, which can be solved in Laplace transform. We call $E$
the conjugate variable of $t$. From
$\tilde{C}(E)$, we get $\tilde{C}(t)$ and then finally compute $C(x,t)$.

\bfig[p]
\bc
\epsfysize=8cm
\epsfbox{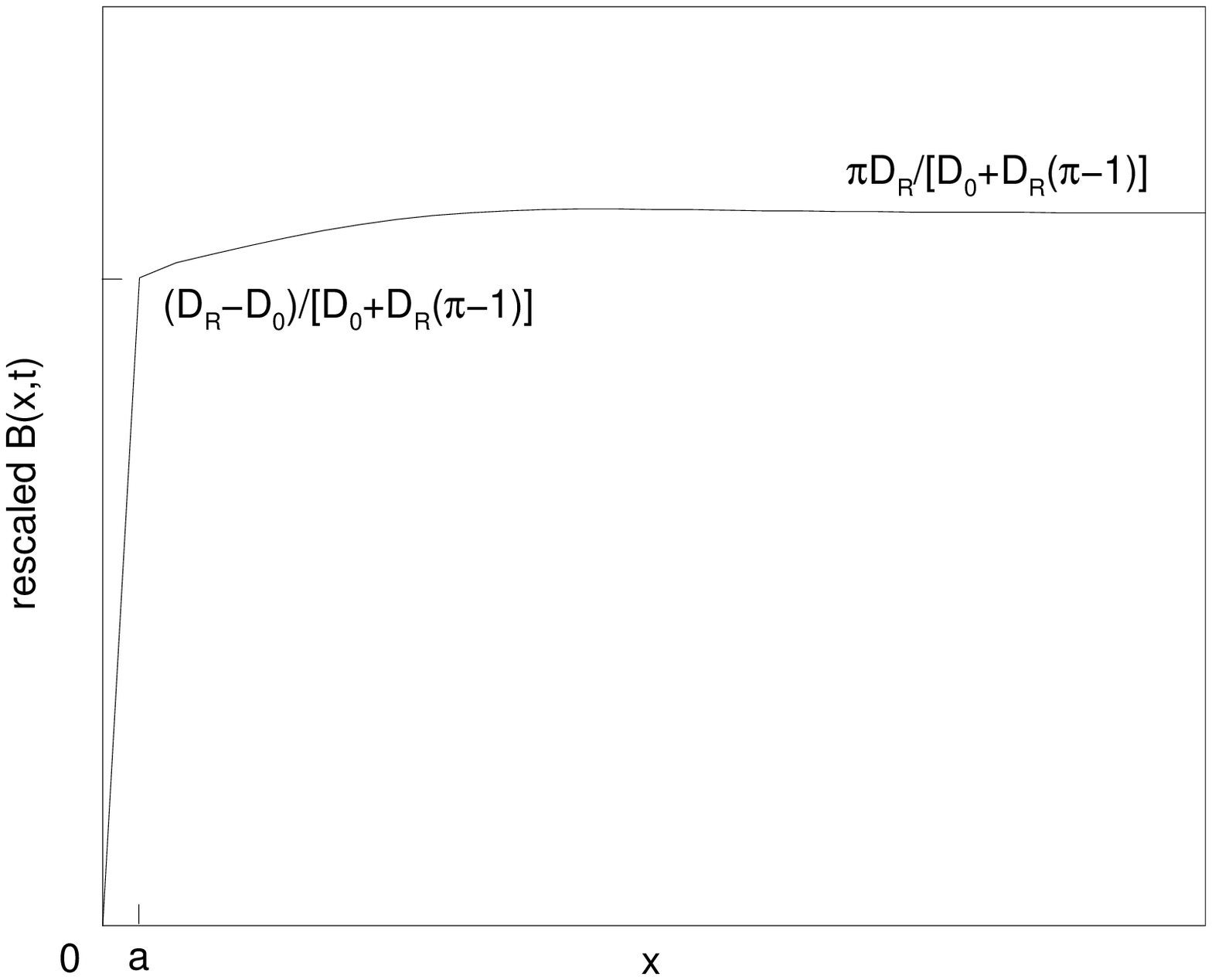}
\caption{\small Correlation function for the case of a random overload. $B$ has been rescaled
by the factor $A_0^2 \over \left [ 8\pi D_R t \right ]^{1/2}$.
\label{CFCfig}}
\ec
\efig
\bfig[p]
\bc
\epsfysize=8cm
\epsfbox{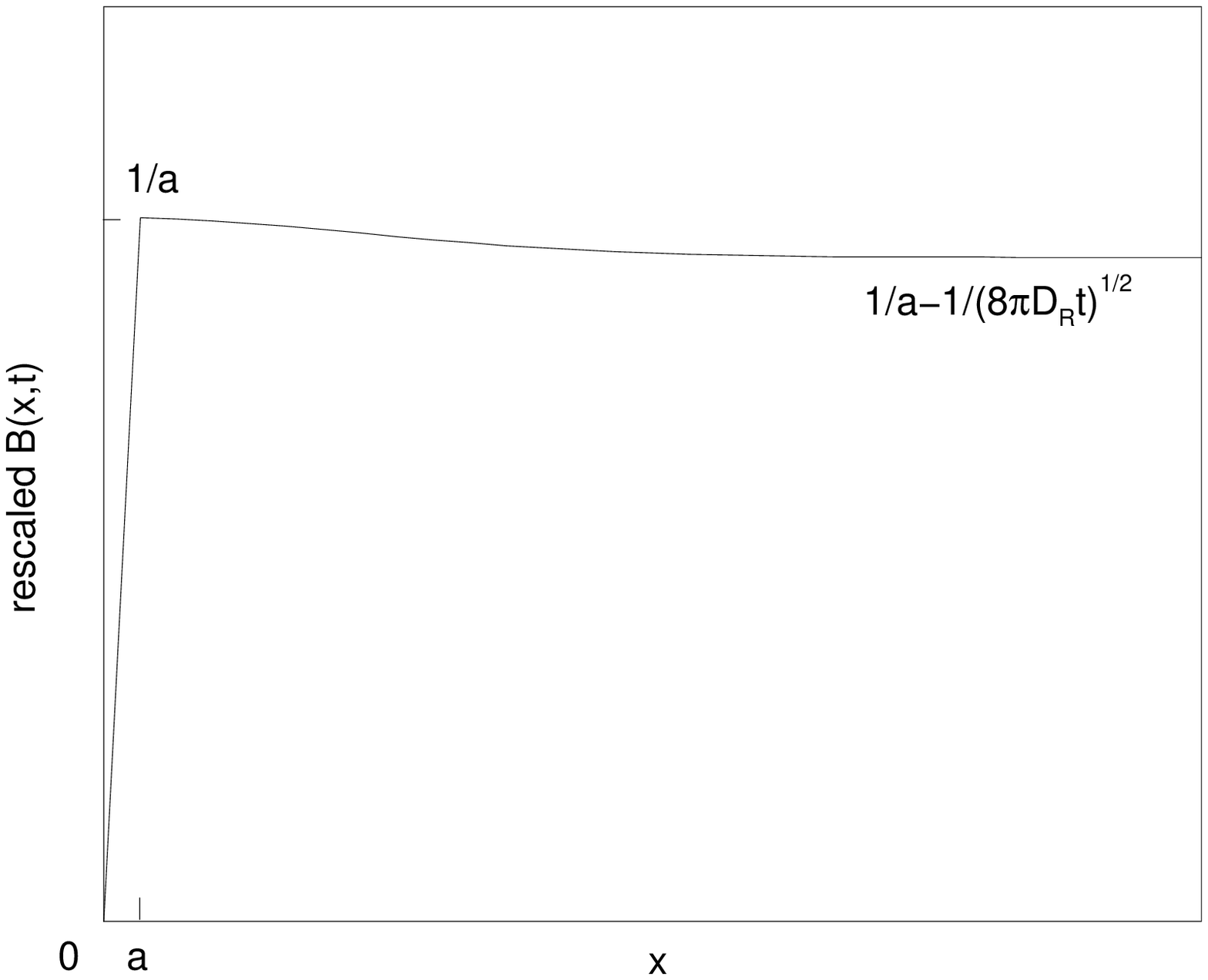}
\caption{\small Correlation function for the case of a uniform overload. $B$ has been rescaled
by the factor $\sigma^2 B_0^2 \over 4[D_0+D_R(\pi-1)]$.
\label{CSCfig}}
\ec
\efig

$\circ$ Random overload: in the small $E$ (large $t$) limit, we get
$\tilde{C}(E) \sim 1/\sqrt{E}$, meaning $\tilde{C}(t) = a_0 /\sqrt{t}$,
with $a_0={\pi D_R \over D_0+D_R(\pi-1)}{A_0 \over \sqrt{8\pi D_R}}$.
It finally leads to the following expression for
$B(x,t) \equiv C(0,t)-C(x,t)=1/2 \la [w(x,t)-w(0,t)]^2 \ra$:
\bea
B(x = 0,t) & = & 0 \nonumber \\
B(x \gg a,t) & = & {A_0^2 \over \sqrt{8\pi D_R t}}
\left [
1-e^{-{x^2 \over 8D_R t}}+{\pi \sigma^2 \over 2[D_0+D_R(\pi-1)]}
\left ({1 \over a} + {x \over 8 D_R t} e^{-{x^2 \over 8D_R t}}
\right ) \right ]
\label{CFC}
\eea
which is shown in figure \ref{CFCfig}.

$\circ$ Constant overload: in the same limit, we get
$\tilde{C}(E) \sim 1/E$, or $\tilde{C}(t) = b_0$,
where $b_0=B_0^2 \ {D_R \over 2[D_0+D_R(\pi-1)]}$. Hence:
\bea
B(x = 0,t) & = & 0 \nonumber \\
B(x \ne 0,t) & = & {\sigma^2 B_0^2 \over 4[D_0+D_R(\pi-1)]}
\left [ {1 \over a} - {1-e^{-{x^2 \over 8D_R t}} \over \sqrt{8\pi D_R t}}
\right ]
\label{CSC}
\eea
which has a form similar to that above: see figure \ref{CSCfig}.

One could have performed the calculation with the Ito convention (corresponding to the Chicago model).
The final results for the correlation function are actually very similar to those above. The main point
is that the correlation is rather structureless. Equation (\ref{CSC}) shows that the correlation function
$C(x > a,t)$ becomes zero for large times, a result that was used to establish equation (\ref{recu}).

\vskip 0.5cm
$\bullet$ Pertubation theory.
\vskip 0.5cm
The above method gives exact results, essentially because $v(x,t)$ is short range
correlated in time: $\delta w/\delta v$ is then only needed at coinciding times, where
it is exactly known. This would not be true in general; furthermore Novikov's
theorem requires $v$ to be gaussian. It is thus interesting to show how a systematic
perturbation scheme can be made to work by the use of diagrams
to represent equation (\ref{liuTF}). The {\sc{mca}} (Mode Coupling Approximation) is
then a particular resummation scheme of this set of diagrams, which was discussed
in detail in \cite{MCA}, which sometimes provide interesting non perturbative results.

Equation (\ref{liuTF}) is multiplied on the left by the operator $G_0$
(see equation (\ref{G01})), and then reexpressed as follows:
\be
\label{mc1}
w(k,t)=G_0(k,t) \otimes \rho (k,t) -i k G_0(k,t) \otimes \int {dq \over 2\pi}
 w(q,t)v(k-q,t)
\ee
$\otimes$ meaning a $t$-convolution product.
This equation can be represented with diagrams as follows: as shown in 
figure \ref{diagdef}, we represent the source $\rho$ by a cross, the `bare'
propagator $G_0$ by a plain line and the noise $v$ by a dashed line. 
\bfig[htb]
\bc
\epsfysize=4cm
\epsfbox{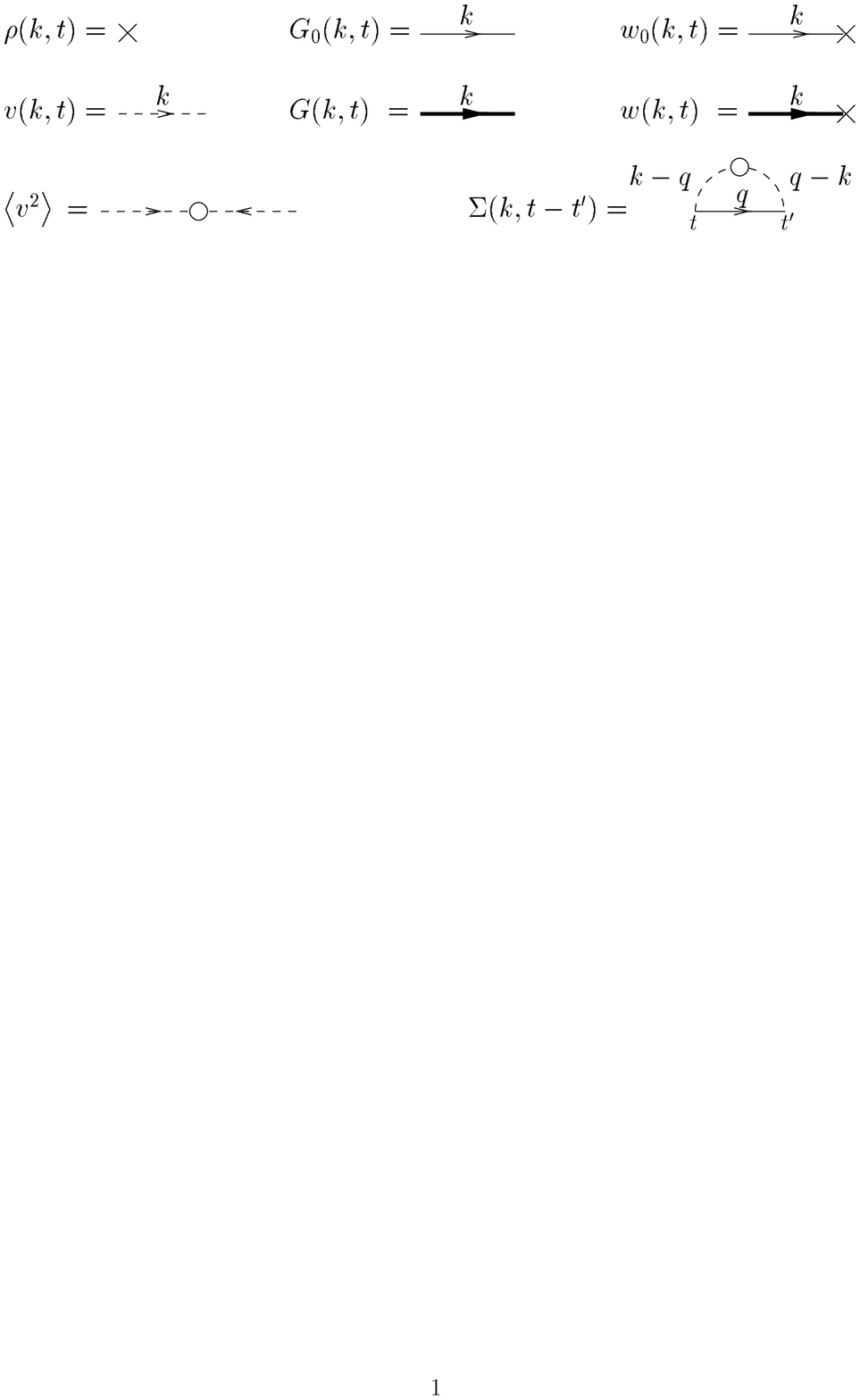}
\caption{\small definition of various diagrams.
\label{diagdef}}
\ec
\efig
The first term of equation (\ref{mc1}), which is the
noiseless solution $w_0$, is then obtained the juxtaposition of a plain line and
a cross. The arrow flows against time (i.e. it is directed from $t$ to $t' < t$). The juxtaposition of
two objects means a $t$-convolution product. By definition $w$ is represented by the
juxtaposition of a bold line and a cross (this is consistent with the identification
of a bold line with the full propagator $G$). The diagrammatic version
of equation (\ref{mc1}) is then:
\be
\label{diagequa1}
\includegraphics{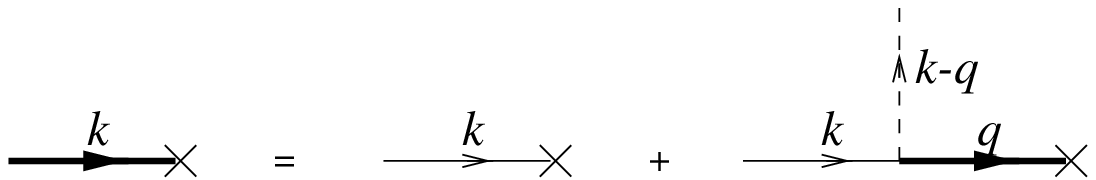}
\ee
The `vertex' stands for $-i k \int {dq \over 2\pi}$, the two emerging
wave vectors being $q$ and $k-q$ (node law). One can now iterate this equation.
To second order, one obtains:
\be
\label{diagequa2}
\includegraphics{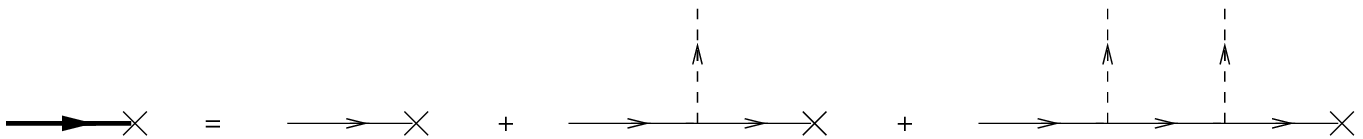}
\ee
The corresponding equation for $G$ is obtained by taking the derivative
$\de / \de \rho$, and averaging over the noise $v$. Since $\la v \ra =0$, the
second diagram vanishes. We represent the noise correlator by
a dashed line with a centered circle (see figure \ref{diagdef}), and 
obtain:
\be
\label{diagequa3}
\includegraphics{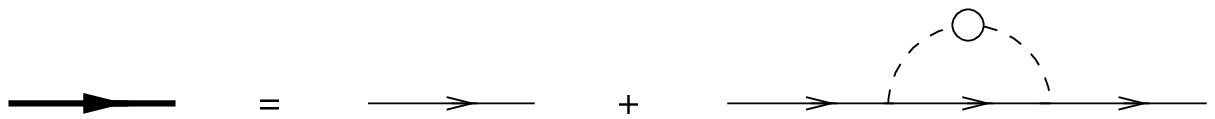}
\ee
or $G = G_0 + G_0 \Sigma G_0$, where $\Sigma$ is called the self-energy
(see figure \ref{diagdef}).  Actually, one can resum exactly all the diagrams
corresponding to $G_0 \Sigma G_0$, $G_0 \Sigma G_0 \Sigma G_0$ to obtain 
the Dyson equation $G = G_0 + G_0 \Sigma G$. 

The {\sc{mca}} amounts to replacing the `bare' propagator in the diagram for $\Sigma$ by the full
propagator $G$.  (Note that the {\sc mca} is of course exact to second order in perturbation 
theory). We then
obtain a self-consistent equation for $G$:
\be
\label{diagequa4}
\Sigma_{{\sc{mca}}} = G_0^{-1} - G_{{\sc{mca}}}^{-1} = 
\includegraphics{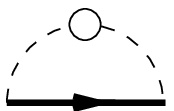}
\ee
Diagrams like the one drawn in figure \ref{diagequa5} are now also
included.
\bfig[htb]
\bc
\epsfysize=2cm
\epsfbox{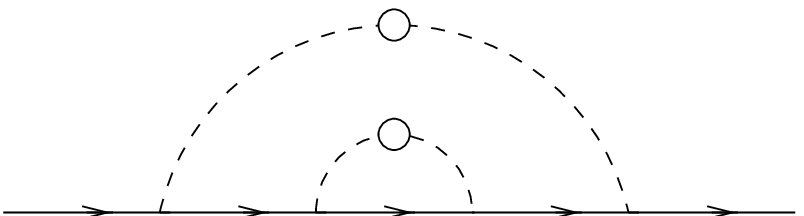}
\caption{\small Example of a diagram included in the {\sc{mca}}.
\label{diagequa5}}
\ec
\efig
The self-energy $\Sigma_{{\sc{mca}}}$ can be easily computed, we get
\be
\label{sigma}
\Sigma_{{\sc{mca}}} (k,t-t') = -2 \pi \sigma^2 k \int {dq \over 2\pi} q G_{{\sc{mca}}}(q,t-t') g_t(t-t')
\ee
In the special case where $g_t$ is peaked around $t=t'$, we can make the approximation
$G(q,t-t') g_t(t-t') \simeq G(q,0) g_t(t-t') = g_t(t-t')$ (since
by definition $G(q,0)=1$).  We thus get, using equation (\ref{convo_q}),
$\Sigma_{\sc{mca}}(k,t-t')= -\sigma^2 \La k^2 g_t(t-t')$.
The expression for $G_{{\sc{mca}}}^{-1}$ is thus identical to the one obtained
with the exact approach, as can be seen by comparing equation (\ref{novi4}) and
$G_0^{-1}G= 1 + \Sigma G$.

Note that one can also calculate the influence of a non
zero kurtosis $\kappa$ of the noise $v$, which is its normalized fourth cumulant.
In this case, four dashed lines (corresponding to $v$) can be merged, leading to a contribution to $D$, of the order of $\kappa
\sigma^4$. 

Let us turn now to the calculation of the correlation function
$\la w(k,t)w(k',t) \ra \equiv 2\pi \de (k+k') C(k,t)$. The basic object which corresponds to 
the self-energy is now the `renormalized source' spectrum $S(k,t,t')$ defined as:
$C=G \otimes S \otimes G$. The quantity $S$ is drawn as a filled square.
$S_0$ (empty square) is the correlation function source
term which encodes the initial conditions (see below).
The two first terms of the expansion are
\be
\label{diagequa6} 
\includegraphics{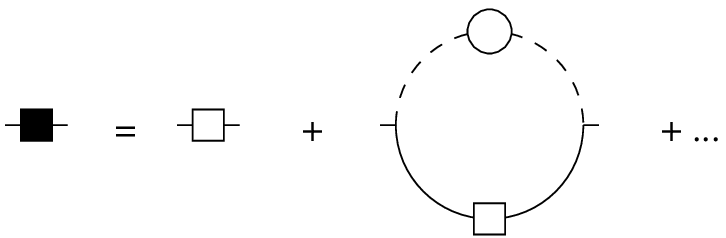}
\ee
Here again, we transform the perturbative expansion into a closed
self-consistent equation for $S$ by replacing $G_0$ and $S_0$ in
(\ref{diagequa6}) by $G$ and $S$ respectively. The final equation for $C$ reads:
\be
\label{diagequa7} 
\includegraphics{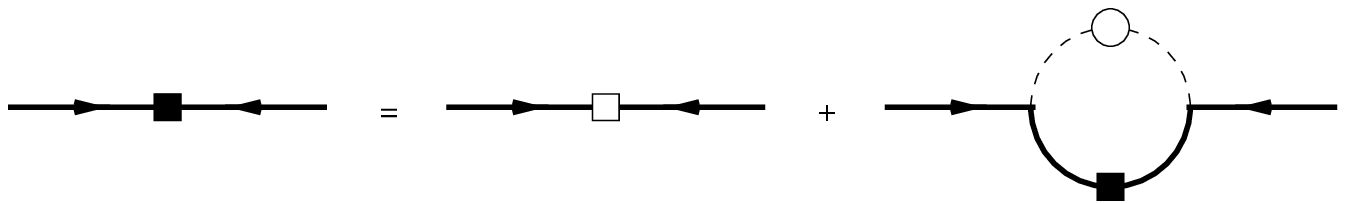}
\ee
or, written explicitly,
\bea
\lefteqn{ C(k,t) = \int_0^t dt' \int_0^t dt''
G(k,t-t')  S_0(k,t',t'')  G(-k,t-t'') + } \nonumber \\
\label{Cmca}
& & \sigma^2 k^2 \int_0^t dt' \int_0^t dt''
G(k,t-t') \int {dq \over 2\pi} C(q,t',t'') g_t(t'-t'') G(-k,t-t'')
\eea
If we choose the source term to be an overload localised at $t=0$, we get:
$S_0 = \la \rho(k,t') \rho(-k,t'') \ra = C(k,0) \de(t') \de(t'')$.

Using the fact that $g_t$ is peaked around $t'=t''$, we again recover
exactly the equation (\ref{C1}) above, showing again that {\sc mca} is 
exact in this special case.

\subsection{Further results: the unaveraged response function}

The {\it average} Green function described above is thus a Gaussian of zero mean, 
and of width growing as $\sqrt{D_R t}$. However, for a {\it given} environment, 
the Green function is not Gaussian, presenting sample dependent peaks 
(see figure \ref{repscalar}). Note however that, contrarily to what we shall find
below for the tensorial case, the unaveraged Green function remains everywhere positive.
\bfig[hbt]
\bc
\epsfysize=8cm
\epsfbox{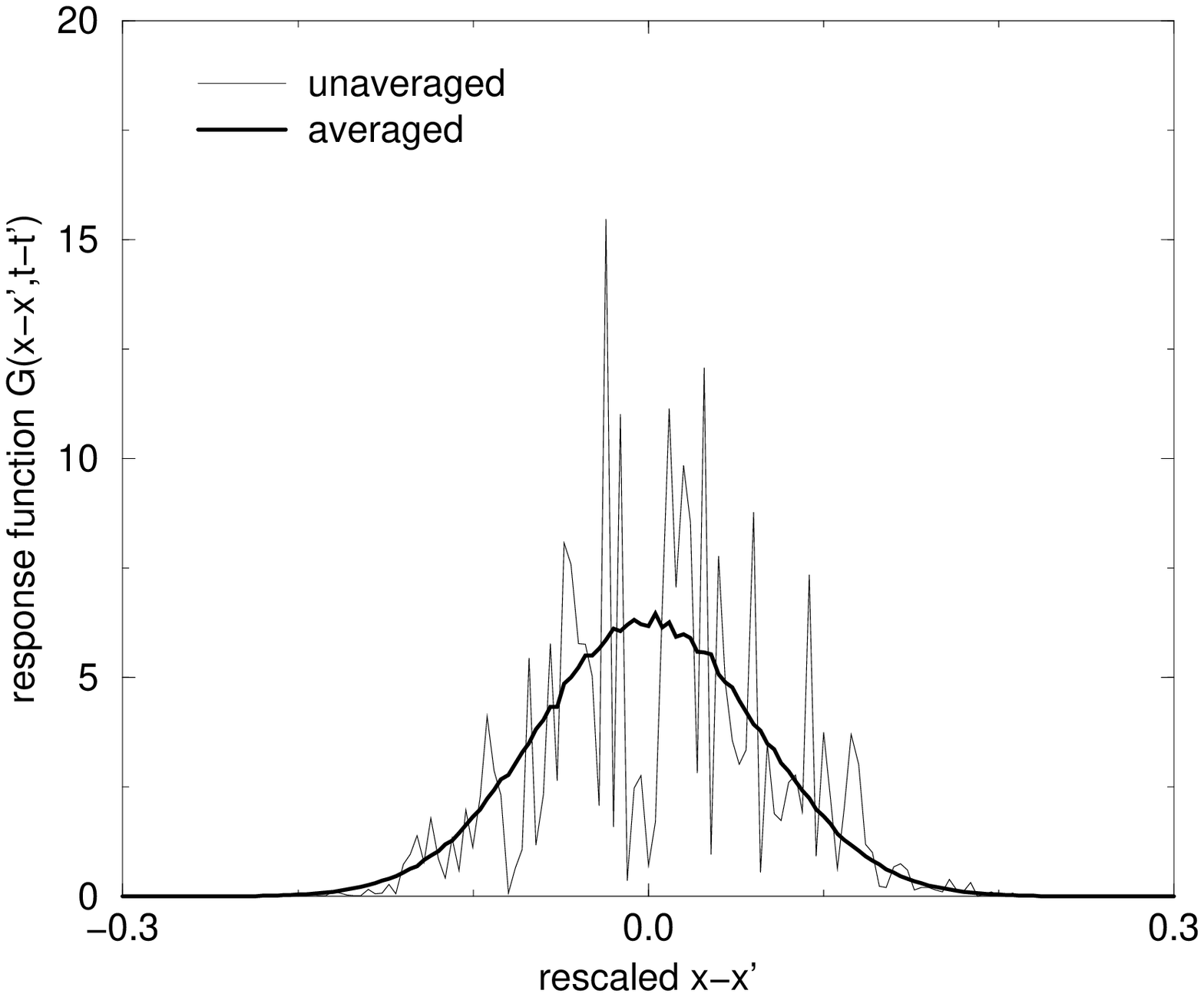}
\caption{\small Averaged (bold line) and unaveraged (thin line) response
function of the scalar model, obtained numerically by simulating
the Chicago model. The average is performed over $5000$ samples.
One can notice how `non self averaging' is the response function, i.e.
how different it is for a given environment as compared to the average. Note
also that the unaveraged Green function is everywhere positive. \label{repscalar}}
\ec \efig Furthermore, the quantity $[x](t)$, defined as the displacement of the
centroid of the weight distribution beneath a point source in a given realisation:
\be
[x](t) = \int_{-\infty}^{+\infty} dx' \  x' \ \frac{\delta w(x',t)}{\delta \rho(0,0)} 
\ee
typically grows with $t$. More precisely, one can show that:
\be
\la [x](t) \ra =0 \quad {\rm but} \quad \la [x]^2(t) \ra \propto t^{1/2}
\ee
meaning that the `center' of Green function wanders away from the origin in a subdiffusive fashion,
as $t^{1/4}$. This behaviour has actually been obtained in an another context, that of a
quantum particle interacting with a time dependent 
random environment. Physically, the Chicago model can indeed be seen as a collection
of time dependent scatterers, converting ingoing waves into outgoing waves with a certain partition
factor $q_+=1-q_-$ (see the discussion in \cite{DirectedWaves}). In two dimensions (plus time), the wandering
of the packet center $[x](t)$ is only logarithmic (and disappears in higher dimensions \cite{DirectedWaves}).

\subsection{The scalar model with bias: Edwards' picture of arches}

Up to now, we have considered the mean value of $v$ to be zero, which reflects the
fact that there is no preferred direction for stress propagation. In some
cases however, this may not be true. Consider for example a sandpile built from a point source:
the history of the grains will certainly inprint a certain oriented `texture' to the contact
network, which can be modelled, within the present scalar model, as a non zero value of
$\la v \ra$, the sign of which depends on which side of the pile is chosen. Let us call
$V_0$ the average value of $v$ on the $x\ge 0$ side of the pile, and $-V_0$ on the other
side. The differential equation describing propagation now reads, in the absence of disorder:
\be
\label{epe}
\partial_t w + \partial_x \left [V_0 \ \mbox{sign}(x)w \right] = \rho + 
D_0 \partial_{xx} w
\ee
\bfig[hbt]
\bc
\epsfysize=8cm
\epsfbox{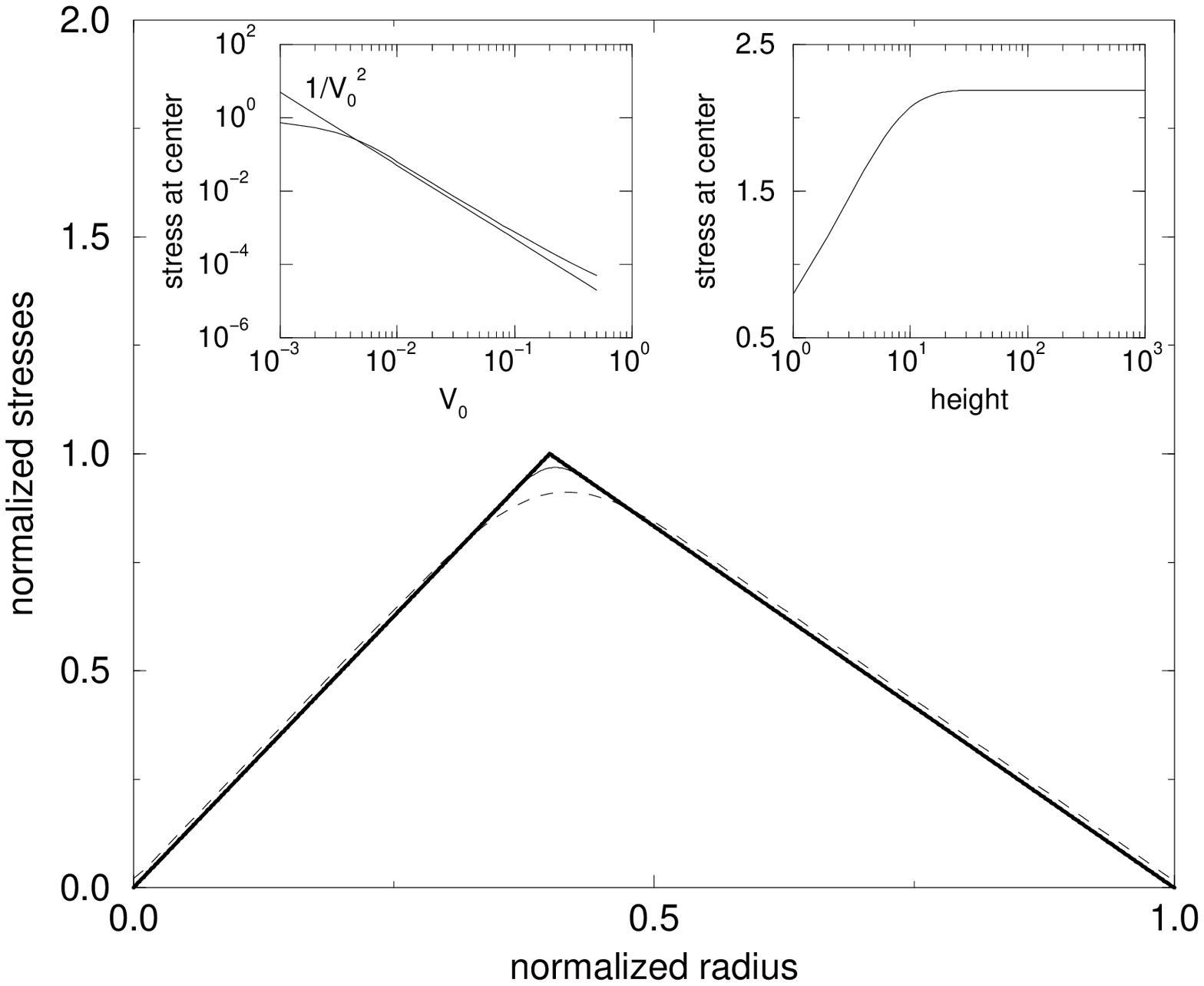}
\caption{\small On the main graph is plotted the solution of equation (\ref{epe}) for $V_0/c=0.4$.
The dashed line is for a diffusion constant $D_0$ ten times smaller than the the
solid one. The bold line is for $D_0=0$. Stress values
are rescaled by the height of the pile $t$. The left inset graph shows
that $w(0,t)$ scales like $1/V_0^2$ at small $V_0$ while the right inset shows that $w(0,t)$ is 
constant for large $t$. Note that for very small values of $V_0$, the $1/V_0^2$ scaling
becomes invalid for finite size reasons.
\label{pep}}
\ec
\efig
(An extra noise can be handled as above). For a constant density $\rho=\rho_0$, and for $D_0=0$,
the weight distribution is then the following: 
\bea
\label{eps}
w(x,t) = & {\rho_0x \over V_0}        & \qquad   \mbox{ for  $0 \le x \le V_0t $} \nonumber\\
w(x,t) = & {\rho_0(ct-x) \over c-V_0} & \qquad  \mbox{ for  $V_0 t\le x \le ct $}
\eea
where $c=1/\tan (\phi)$ ($\phi$ is the angle made by the slope of the
pile with the horizontal $x$ axis). For $D_0 \neq 0$, the above solution is smoothed
(see figure \ref{pep}).
In any case, the local weight reaches a {\it minimum} around $x=0$.  Equation (\ref{epe}) gives a
precise mathematical content to Edwards' model of arching in sandpiles \cite{Edwards},
as the physical mechanism leading to a `dip' in the pressure distribution \cite{Smid}.
As discussed elsewhere \cite{WCCB,FPA}, this can be taken much further within a
tensorial framework (see section \ref{tensorial}). 

The scaling of the stress at the center of the pile, $w(0,t)$, can be understood simply in terms of
random walks subject to a bias $V_0$. The region contributing to $w(0,t)$ is then found to be of
finite volume, independent of $t$, and of the order of $D_0/V_0^2$, as shown on the two top
pictures of figure \ref{pep}. 

Equation (\ref{epe}) with noise can in fact be obtained naturally within an extended Chicago model,
with an extra rule accounting for the fact that a grain can slide and lose contact with 
one of its two downward neighbours \cite{CB}. This generically leads to arching; in the sandpile
geometry and for above a certain probability of (local) sliding, the effective `velocity' $V_0$
becomes non zero and the weight profile (\ref{eps}) is recovered \cite{CB}. However, this extra
sliding rule implicitly refers to the existence of shear stresses, which are absent in the scalar
model, but which are crucial to obtain symmetry breaking effects modelled by a non zero $V_0$.
It is thus important to consider from the start the fact that stress has a tensorial,
rather than scalar, nature. This is what we investigate in the following section.

\section{The Tensorial Model}
\label{tensorial}

\subsection{The wave equation}

It is useful to start with a simple `toy' model for stress propagation, which is the analogue 
of the model presented in figure \ref{liufig}. We now consider the case of three
downward neighbours (see figure \ref{threeleg}), for a reason which will become clear below.
\bfig[hbt]
\bc
\epsfysize=4cm
\epsfbox{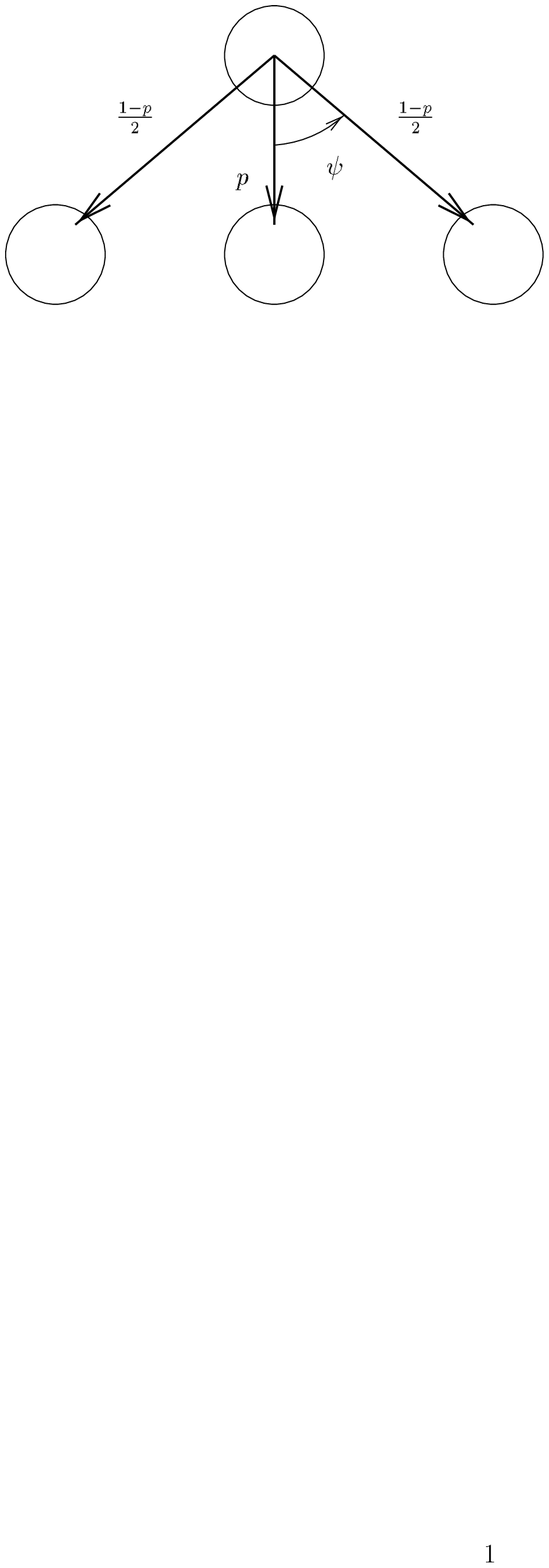}
\caption{\small Three neighbour configuration. Each grain transmits two force components
to its downward neighbours. A fraction $p$ of the vertical component is
transmitted through the middle leg, and a fraction $(1-p)/2$ through each of the external legs.
\label{threeleg}}
\ec
\efig
Each grain transmits to its downward neighbours not one, but two force components: one along
the vertical axis $t$ and one along $x$, which we call respectively $F_t(i,j)$ and $F_x(i,j)$.
(We will restrict attention, in the sequel, to two-dimensional piles, leaving extensions to
three dimensions for further investigations.) For simplicity, we assume that the `legs'
emerging from a given grain can only transport 
the vector component of the force parallel to itself (but more general rules could be invented). Assuming
that the transmission rules are locally symmetric, and that a fraction $p \leq 1$ of the 
vertical component travels through the middle leg, we find:
\bea
F_x(i,j) & = & \frac{1}{2} \left[F_x(i-1,j-1)+F_x(i+1,j-1)\right] \nonumber \\
         &   & + \frac{1}{2} (1-p) \tan \psi \left[F_t(i-1,j-1)-F_t(i+1,j-1)\right] \label{wemicro1}\\
F_t(i,j) & = & w_0 + p F_t(i,j-1) + \frac{1}{2}(1-p) \left[F_t(i-1,j-1)+F_t(i+1,j-1)\right] \nonumber \\
         &   & +  \frac{1}{2 \tan \psi} \left[F_x(i-1,j-1)-F_x(i+1,j-1)\right] \label{wemicro2}
\eea
where $\psi$ is the angle between grains, defined in figure \ref{threeleg}. Taking 
the continuum limit of the above equations leads to:
\bea
\label{equiF1}
\partial_t F_t + \partial_x F_x & = & \rho \\
\label{equiF2}
\partial_t F_x + \partial_x \left [ c_0^2 F_t \right ] & = & 0
\eea
where $c_0^2 \equiv (1-p) \tan^2 \psi$. Eliminating (say) $F_x$ between the above two equations
leads to a {\it wave equation} for $F_t$, where 
the vertical coordinate $t$ plays the r\^ole of time and 
$c_0$ is the equivalent of the `speed of light'. In particular, the
stress does not propagate vertically, as it does in the scalar model, but rather at a
{\it non zero angle} $\varphi$ such that $c_0=\tan\varphi$. Note that $\varphi \neq \psi$
in general (unless $p=0$); the angle at which stress propagates has nothing to do with 
the underlying lattice structure and can in principle be arbitrary. We chose
a three leg model to illustrate this particular point.

The above derivation can be reformulated in terms of classical continuum mechanics
as follows. Considering all stress tensor components $\sigma_{ij}$, the equilibrium equation reads, 
\bea
\label{equi1}
\partial_t \szz + \partial_x \sxz & = & \rho \\
\label{equi2}
\partial_t \szx + \partial_x \sxx & = & 0
\eea
Identifying the local average of $F_t$ with $\szz$ and that of $F_x$ with $\szx$, 
we see that the above equations (\ref{equiF1}, \ref{equiF2}) are actually identical to 
(\ref{equi1}, \ref{equi2}) provided $\szx=\sxz$ (which corresponds to the absence of local torque)
and $\sxx = c_0^2 \szz$. This relation between normal stresses was postulated
in \cite{BCC} as the simplest constitutive relation obeying the correct symmetries 
which enables one to lift the indeterminacy of equations (\ref{equi1}, \ref{equi2}); 
it can be seen as a {\it local} Janssen approximation \cite{Janssen}.
$c_0^2$ should encode the relevant information of the local geometry of the packing, 
friction, shape of grains, etc., and should thus depend on the construction 
history of the grain assembly. For example, in the sandpile geometry $c_0^2$ is related 
to the angle of friction $\phi$ of the material by the relation $c_0^2=1/(1+2\tan^2\phi)$ \cite{BCC}.
This approach can be generalized to take into account a
local asymmetry in the packing texture (which one expects for example in the case
of a sandpile constructed from a point source), by allowing $c_0^2$
to depend on $\sxz/\szz$ \cite{BCC,WCCB,FPA}. If this dependence is linear, this is equivalent to a coordinate rotation in $x,t$ \cite{FPA}.
 
\subsection{A stochastic wave equation}

The starting point of the scalar model is thus essentially the diffusion equation,
which one perturbs by adding a random convective term. As the above paragraph shows, a more 
natural starting point is the wave equation. The toy model presented above however
suggests that, provided local conservation laws are obeyed (i.e. those arising from
mechanical equilibrium), many local rules for force transmission are compatible with
the contact conditions \cite{Eloy}. It is thus natural to encode the disorder of
the packing, or model the indeterminacy of the contact conditions
as a randomly varying `speed of light' $c_0$ (reflecting the fact that, for example, the
parameter $p$ can vary from grain to grain).
\bfig[hbt]
\bc
\epsfysize=8cm
\epsfbox{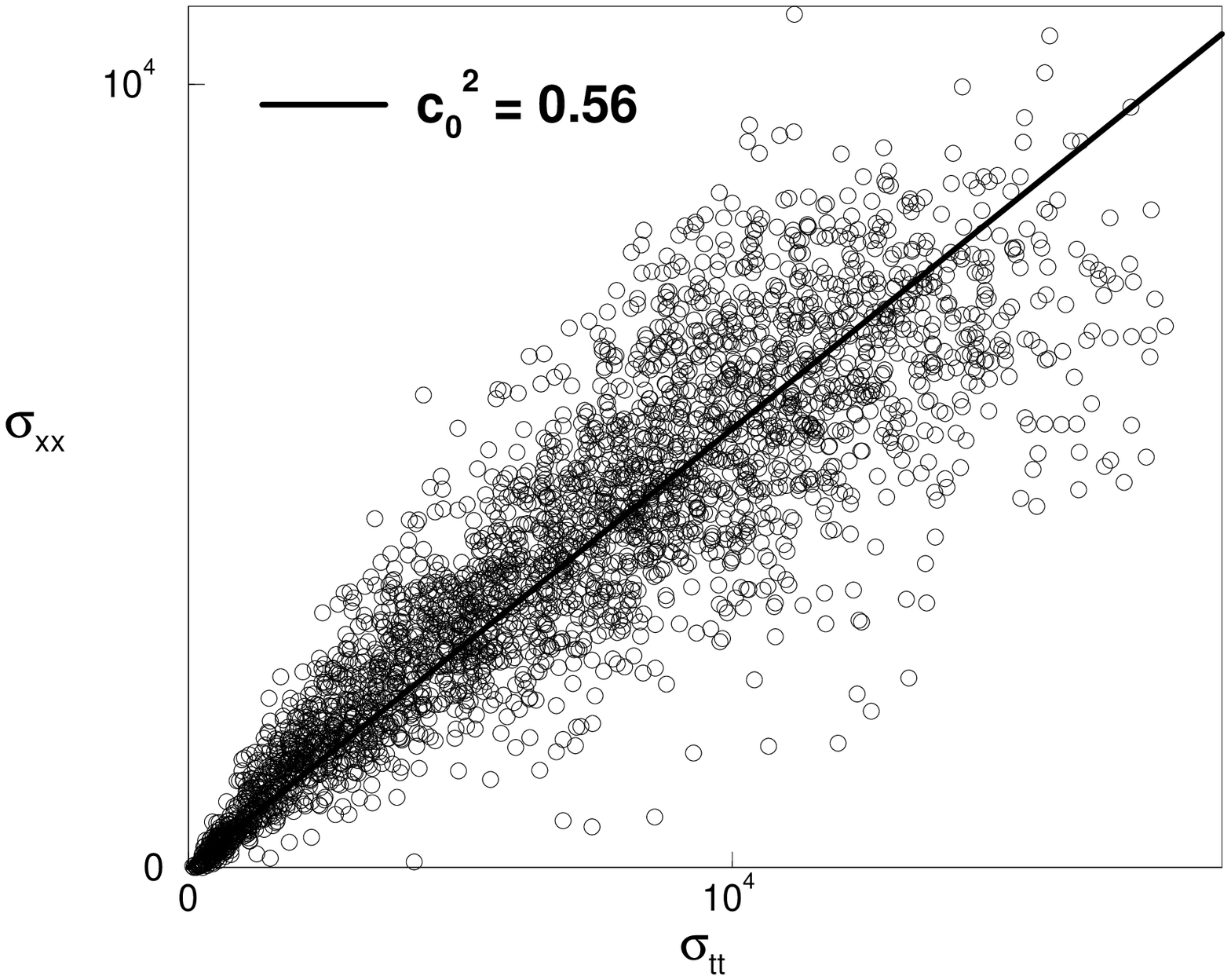}
\caption{\small Relation between $\sxx$ and $\szz$ from a microscopic numerical simulation
of grains forming a heap in two dimensions \protect\cite{Jojo}.
These data are compatible with a stochastic constitutive relation
$\sxx =c_0^2 [1+v(x,t)]\szz$, where $v$ is a random noise.
\label{Jojo1}}
\ec
\efig
Two recent numerical simulations \cite{Eloy,Jojo}
actually suggest that this should be a good first approximation. In figure \ref{Jojo1},
we show a scatter plot of $\sxx$ versus $\szz$, measured as averages of the local forces
over a small box centered around different points within a heap (from
ref.\cite{Jojo}).
This plot clearly shows that a linear relation is indeed acceptable, leading
in this case to $c_0^2 \sim 0.56 \pm 0.03$ \cite{Jojo}.
There are however significant fluctuations, reflecting some disorder in the packing,
which are furthermore uncorrelated from point to point. The histogram of $v$ defined as: 
\be
\sxx = c_0^2 [1+v(x,t)] \szz \qquad \la v(x,t) \ra =0
\label{CRR}
\ee
is found to be roughly gaussian, of relative width $\sigma \sim 0.3$. This corresponds to a locally
varying angle of stress propagation, which varies around the mean angle $54^o$ 
by an amount $\sim 10.8^o$. 

Motivated by the simulations results, we now investigate a model
(called `random symmetric model' in the following)
with the inhomogeneous constitutive relation (\ref{CRR}), which leads to the
following stochastic wave equation for stress propagation:
\be
\partial_{tt} \szz = \partial_{xx} \left[c_0^2(1+v(x,t))\ \szz\right]\label{RW}
\ee
where the random noise $v$ is assumed to be correlated as 
$\la v(x,t)v(x',t') \ra=\sigma^2 g_x(x-x') g_t(t-t')$. The correlation lengths $\ell_x$ and $\ell_t$
are again kept finite, and of the same order of magnitude.
In Fourier transform, this relation can also be written
$\la v(k,t)v(k',t') \ra=2\pi \sigma^2 \de(k+k') \tilde g_x(k) g_t(t-t')$. It
turns out that the final shape of the averaged response function depends on the
sign of $\tilde g_x(\La)$. In section \ref{Scalar} we implicitly made the
choice $\tilde g_x(k)=1$, which corresponds to:
$g_x(x=0)=1/a$ and $g_x(x > 0)=0$. We will keep this choice for the following
calculations, but note that another form for $g_x$ could lead to
$\mbox{sign}(\tilde g_x(\La))=- 1$.  

In the following, $\szz$ will be again denoted by $w$. After a Fourier transform along $x$-axis,
we get, from equation (\ref{RW})
\be
\label{szz}
(\partial_{tt} + c_0^2k^2) w = \partial_t \rho -
c_0^2k^2 \int {dq \over 2\pi} w (q,t) v(k-q,t)
\ee
Note that the `source' term of this equation is now $\partial_t \rho$ rather
than $\rho$ itself. Therefore,
if we call $G$ the Green function (or propagator) of this equation
$G=\la \de w / \de \partial_t \rho \ra$; the response function
$R=\la \de w / \de \rho \ra$ of our system is now actually the
time derivative of $G$: $R(k,t)=\partial_t G(k,t)$.

The noiseless propagator $G_0$ is the solution of the ordinary wave equation
$(\partial_{tt}+c_0^2k^2)G_0(k,t-t')=\de (t-t')$ and can be easily calculated:
\be
\label{P0}
G_0(k,t) = {1 \over 2 i c_0 k} \left [e^{i c_0 k t}
- e^{-i c_0 k t} \right ] \theta (t)
\ee
which leads to the response function $R_0$
\be
\label{G0vectreal}
R_0(x,t) = {1 \over 2}
\left [ \de \left ( x-c_0 t \right ) + 
\de \left ( x+c_0 t \right ) \right ] \theta (t)
\ee
This last equation sums up one of the major results of \cite{BCC} (see also \cite{WCCB,FPA}):
in two dimensions, stress propagates
along two characteristic rays. (Note that the corresponding response function in three dimensions
\footnote{In three dimensions a secondary closure is needed, for instance
$\sigma_{xx} = \sigma_{yy}$, $y$ being the third coordinate. \label{3D} }
reads $R_0(x,t) \propto (c_0^2 t^2 -x^2)^{-1/2}$ for $|x| < c_0 t$ and zero otherwise \cite{BCC}).
A relevant question is now to ask how these rays survive in the presence of disorder.
We will show that for weak disorder, the $\de$-peaks acquire a finite (diffusive) width,
and that the `speed of light' is renormalized to a lower value. Not surprisingly, the effect
of disorder can be described by an `optical index' $n>1$. For a strong disorder, however, we find (within 
a gaussian approximation for the noise $v$) that the speed of light vanishes and then becomes imaginary.
The `propagative' nature of the stress transmission disappears and the system behaves more like an
elastic body, in a sense clarified below.

\subsection{Calculation of the averaged response function}

One can again use Novikov's theorem in the present case if the noise is gaussian and
short range correlated in time. However, the same results are again obtained within
the diagrammatic approach explained in section \ref{Scalar}, which can be easily
transposed to the present case, and is more general. The propagator $G$ is a now
represented as a line, the source $\partial_t \rho$ a cross and the vertex meaning
$-c_0^2k^2\int{dq \over 2\pi}$. Within the {\sc mca}, the self-consistent equation
(analogous to equation (\ref{diagequa4}) in the scalar case) is:
\be
\label{H}
(\partial_{tt} +c_0^2k^2)H(k,t)=\delta(t) + \int_0^t dt' \Sigma_{\sc mca}(k,t') H(k,t-t')
\ee
where $H$ is defined by $G(k,t)=H(k,t)\theta(t)$, and the self-energy
$\Sigma_{\sc{mca}}$ given as
\be
\label{sevect}
\Sigma_{\sc{mca}} (k,t-t') =
2 \pi c_0^4 \sigma^2 k^2 \int {dq \over 2\pi} q^2 g_t (t-t')\tilde g_x(k-q) H(q,t-t')
\ee
Equation (\ref{H}) can be solved using a standard Laplace transform along
the $t$-axis ($E$ is the Laplace variable). Using the fact that $H(k,\tau)= \tau$ in
the limit where $\tau \to 0$, we find, for small $k,E$ (corresponding to scales $L$ such
that $\ell_x,\: \ell_t \ll L$): $H^{-1}(k,E)=E^2+\beta E+c_R^2k^2$, where
\bea
\label{coeff1}
c_R^2(k) & = & c_0^2 - {c_0^4 \sigma^2 \La^3\ell_t \over 6}\left 
(1-{3|k| \over 2\La} \right )
+ {\cal O}(k^2)\\
\label{coeff2}
\beta(k) & = & {c_0^4 \sigma^2 k^2 \La^3 \ell_t^2 \over 9} + {\cal O}(k^3)
\eea

We notice here that in the limit $\ell_t \to 0$, the effect of the randomness completely
disappears, as in the scalar model with the Ito convention. (Technically, this is due to the
fact that $G(k,t=0) \equiv 0$ in the present problem). In order to calculate the inverse
Laplace transform, we need to know the roots of the equation $H^{-1}(k,E) = 0$. This
leads to several phases, depending on the strength of the disorder. 

\vskip 0.5cm
$\bullet$ The weak disorder limit.
\vskip 0.5cm
For a weak disorder, $c_R^2(k)$ is always positive. We can then define $c_R=c_R(k=0)$.
As we will show now, $c_R$ is the shifted `cone' angle
along which stress propagates asymptotically. $c_R$ is a decreasing function of $\sigma$,
meaning that the peaks of
the response function get closer together as the disorder increases \footnote{As a technical 
remark, let us note that 
if $g_t=g_x$, the problem is symmetric in the change $x \to t$, $c_0^2(x,t) \to 1/c_0^2(x,t)$. 
It thus looks as if the cone should both narrow or widen, depending on the arbitrary choice
of $x$ and $t$. There is however no contradiction with the above calculation since we assumed that 
$v$ has zero mean, while $1/(1+v)-1$ has a positive mean value, of order $\sigma^2$.}. For a critical 
value\footnote{For $\ell_t=\ell_x=1$ and $c_0^2=0.6$ (corresponding to $\phi=30^o$),
one finds $\sigma_c \simeq 0.57$.} $\sigma=\sigma_c$, $c_R$ vanishes, and becomes
imaginary for stronger disorder.

In the limit of large $t$, the propagator reads: \be
\label{Pa}
G(k,t)= {1 \over c_R k} \sin \left [
c_R kt(1+\alpha |k|) \right ] e^{-\ga k^2t} \theta(t)
\ee
where the following constants have been introduced \footnote{Note that the sign of $\alpha$ is dictated by the sign of $\tilde g_x(\Lambda)$.}:
\bea
\label{ctes1}
\alpha &=& {3 \over 4\La} \left ( {c_0^2 \over c_R^2} -1 \right ) \\
\label{ctes2}
\ga &=& {\beta(k) \over 2k^2} = {\sigma^2 \La^3 \ell_t^2 \over 18}
\eea
From equation (\ref{Pa}), the response function $R$, in the limit of small $k$ and large $t$,
is given by:
\be
\label{R}
R(k,t)= \cos \left [
c_R kt(1+\alpha |k|) \right ] e^{-\ga k^2 t} \theta(t)
\ee
or in real space,
\be
\label{Gvw}
R(x,t)  = {1 \over 2\sqrt{4\pi|\gah|(t)}}
\Re \left\{ {e^{-\xi_+^2/ b} \over \sqrt{b}} 
\left [1-\Phi (-i {\xi_+ \over \sqrt{b}}) \right ] +
\sqrt{b}\: e^{-b \xi_-^2}
\left [1-\Phi (-i \sqrt{b} \xi_-) \right ] \right \}
\ee
\bfig[hbt]
\bc
\epsfysize=8cm
\epsfbox{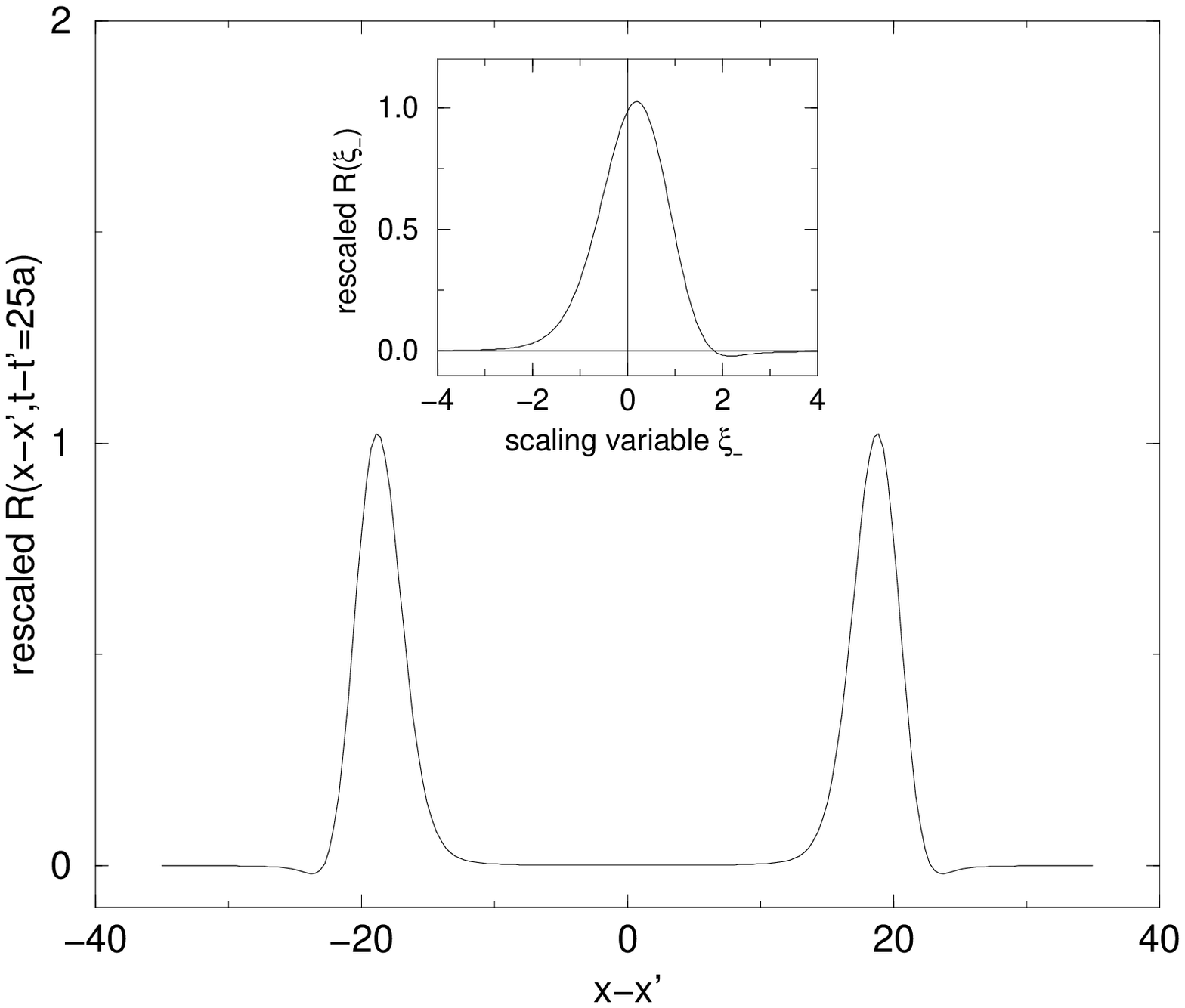}
\caption{\small Response function for weak disorder ($\sigma/\sigma_c \sim 0.32$).
The two curves have been rescaled by
the factor $2\left [4\pi|\gah|t \right]^{1/2}$.
The main graph shows the general double-peaked shape of the
response of the system when subjected to a peaked overload at $x=0,\ t=0$.
The inset gives details the right-hand peak as a function of the scaling
variable $\xi_-$. Note the asymmetry (for $\tilde g_x(\Lambda)>0$), compatible with
the results found in \protect\cite{Eloy}. Note also that the curve becomes
negative around $\xi_-=2$.
\label{Gwdfig}}
\ec
\efig
where the scaling variables $\xi_\pm$, measuring distances relative to the two peaks, 
are defined by
\be
\label{xipm}
\xi_\pm = {x \pm c_R t \over \sqrt{4|\gah|t}}
\ee
and where $\gah = \ga - i c_R \alpha$ and 
$b = e^{i \arg{\hat{\ga}}}$. $\Phi$ is the standard error function.
Figure \ref{Gwdfig} shows $R$ as given by expression (\ref{Gvw}).
Interestingly, this propagator not only has a finite diffusive width $\propto \sqrt{t}$, but is also asymmetric around its maxima.
Surprisingly, and in sharp contrast to the scalar case discussed above, the response function becomes {\it negative} in certain intervals (although its integral is of course
equal to one because of weight conservation). This means that pushing on a given point actually reduces
the downward pressure on certain points. This can be interpreted as some kind of arching: increasing the
shear stress does affect the propagation of the vertical stress, and may indeed 
lead to a
reduction in its local value which is redistributed elsewhere. As we shall see in
section \ref{numerics}, the unaveraged response function indeed takes negative
(and rather large) values. This is a very significant result since it suggests
that granular materials may be susceptible to rearrangement
under extremely weak external perturbations.
Suppose indeed that as a result of the perturbation, a
grain receives to a negative force larger than the preexisting vertical pressure.
This grain will then move and a local rearrangement of contacts will occur,
inducing a variation of $c_0(x,t)$ as to reduce the cause of the instability.
Thus, the stochastic wave
equation implicitly demands rules similar to those introduced in \cite{CB} to
describe extreme sensitivity to external perturbations in silos. The present
model, which is purely static, does not say what to do when a local rearrangement
occurs, but certainly suggests that small perturbations will induce such
rearrangements.  

It is interesting to note that this response function was numerically measured in 
ref. \cite{Eloy}; its
shape is compatible with the above expression; in particular, the two peaks were 
found to be
asymmetric with a longer `tail' extending inwards, as we obtain here.
Note however that for $\tilde g_x(\La)<0$, the shape of the
peaks is reversed: the small dips are located inside the peaks and the longer
tail extends outwards. This is actually what we obtain numerically in section
\ref{numerics}. 

\vskip 0.5cm
$\bullet$ Shear response function.
\vskip 0.5cm
Equation (\ref{equi1}) provides a straightforward way to calculate the shear response
function $R_s$ in terms of $R$. Indeed, one has: $i k R_s(k,t) = \de (t) - \partial_t R(k,t)$.
We thus get, in the limit of small $k$ and large $t$,
\be
\label{R'}
R_s(k,t)=-i c_R \sin \left [ c_R kt(1+\alpha |k|) \right ] e^{-\ga k^2 t} \theta(t)
\ee
This shear response function is very similar to $R$, except that it is, as expected, 
an odd function of $x$.

\vskip 0.5cm
$\bullet$ Effective large scale equations.
\vskip 0.5cm
It is interesting to know of which differential equations the response functions $R$ and $R_s$,
are solutions. These effective
equations can be interpreted as a coarse-grained (hydrodynamical) description of the  
propagation of a stress perturbation which takes into account the average effect of the local disorder.
One problem however comes from the presence of the `dispersion' term  $\alpha |k|$, which corresponds to 
a non-local operator in real space. We thus neglect this term in the following discussion,
but one should keep in mind that the effective equation are actually non local. In any 
case, the main features of the
response functions (peaks centered around $x=\pm c_R t$ ($c_R < c_0$) with a diffusive width
$\propto \sqrt{t}$) are not lost when setting $\alpha=0$ (except for the fact that the response 
function can become negative which is related to $\alpha \neq 0$). Effective equations can then be written 
in the large $t$ limit as:
\bea
\label{effequa1}
\partial_t \la \delta \szz \ra & = & \delta \rho - \partial_x \la \delta \sxz \ra  \\
\label{effequa2}
\partial_t \la \delta \sxz \ra & = & - c_R^2 \partial_x \la \delta \szz \ra + 2\ga \partial_{xx} \la \delta \sxz \ra
\eea
That disorder generates the diffusion terms $2 \ga \partial_{xx} \la \delta\sxz \ra$ is rather
intuitive and had been guessed in \cite{BCC}. This terms can be seen the first term of a
gradient expansion of to the constitutive equations which have the correct symmetry, i.e.:
\bea
\label{cr1}
\la \delta\sxx \ra  & = & c_R^2 \la \delta\szz \ra - \ga \partial_x \la \delta\sxz \ra \\
\label{cr2}
\la \delta\szx \ra & = & \la \delta\sxz \ra 
\eea
where the second equation is imposed by the absence of local torque.  

We have thus shown that the introduction of a small disorder in the local direction of propagation
does not change radically the nature of stress propagation on large length scales, although the
peaks in the response function acquire a diffusive width. These peaks acquire a width of the order of
$\sqrt{\gamma H}$ (where $H$ is the height of the pile), and are thus well separated in the 
limit where $H \gg \gamma$. As we shall see now, this is no longer true if the disorder becomes strong. 

\vskip 0.5cm
$\bullet$ Critical disorder: The wave/diffusion transition.
\vskip 0.5cm
When the disorder is so strong that $c_R$ just vanishes, the roots of
$H^{-1}(k,E) = 0$ change nature, and so does the response function $R$.
The two peaks of the previous expression for $R$ merge together, while the width
becomes anomalously large ($\propto t^{2/3}$). In the asymptotic, large $t$, regime we obtain:
\be
\label{Pvs}
R(k,t) = \theta(t)\cos \left [ \lambda |k|^{3/2} t \right ] e^{-\ga k^2 t}
\ee
where the new constant $\lambda$ is defined by $\lambda=c_0\sqrt{3/2\La}$ and $\ga=c_0^2\ell_t/3$.
\bfig[htb]
\bc
\epsfysize=8cm
\epsfbox{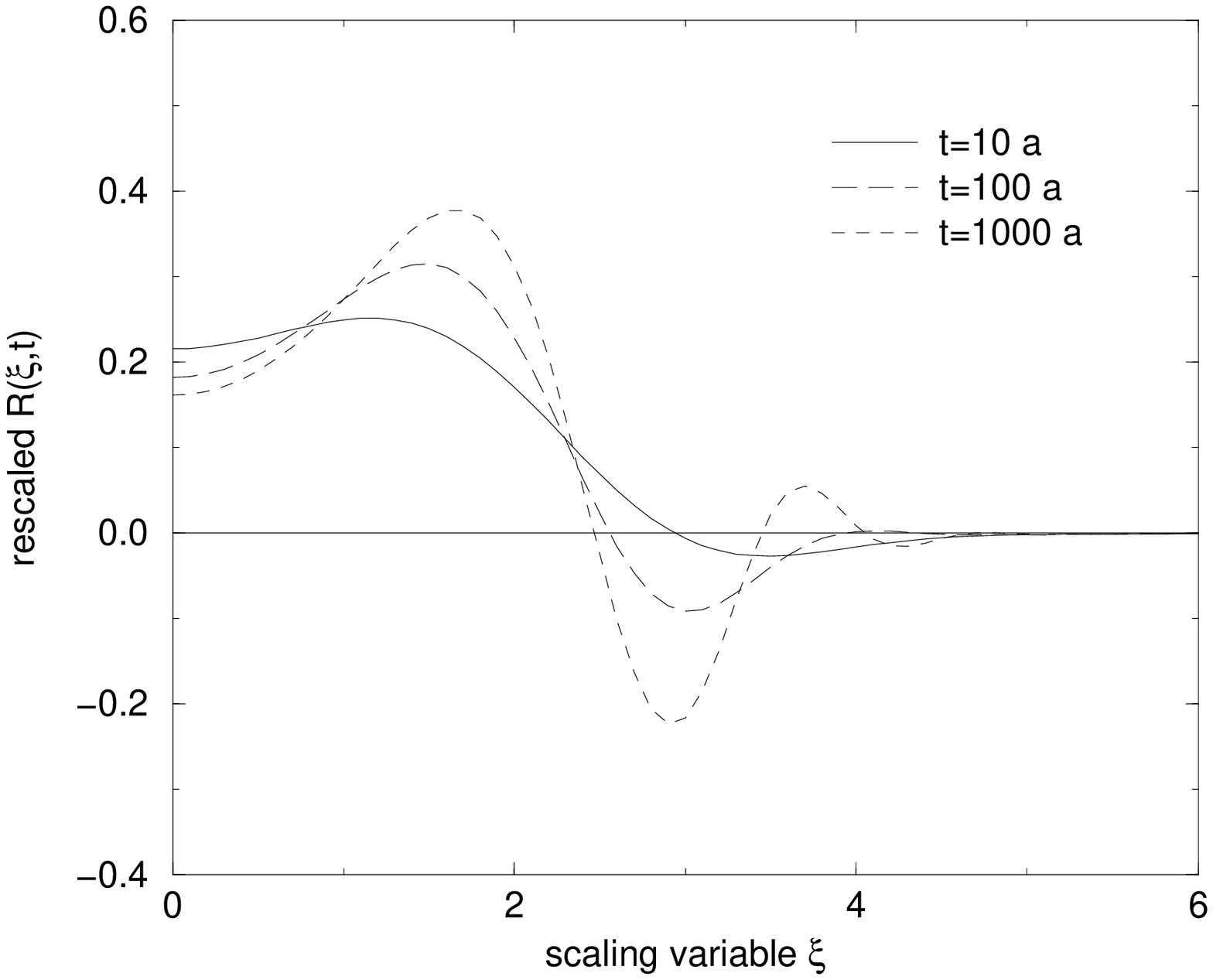}
\caption{\small Response function for a critical disorder: $c_R=0$.
\label{Gsdfig}}
\ec
\efig
The physical response function $R$ is plotted in figure \ref{Gsdfig}, for different 
values of $t$, as a function of the scaling variable  
\be
\label{xi}
\xi = {x \over \lambda t^{2/3}}.
\ee
On the scale $t^{2/3}$, the double peak structure of $R$ is still visible. However, note that 
the term $e-\ga k^2 t$ cannot be neglected, even for large $t$; this means that the response 
function is never really a function of $\xi$ only, as clear from figure \ref{Gsdfig}. Note that
the response function again becomes negative for some values of $\xi$.

\vskip 0.5cm
$\bullet$ Strong disorder: The pseudoelastic regime.
\vskip 0.5cm

For larger disorder still, one finds, within the {\sc mca} (which is exact for a gaussian,
uncorrelated noise), that the renormalized value of $c_0^2$, $c_R^2$, becomes negative.
Upon a rescaling of $x$ as $\hat x = x/(ic_R)$, the effective equation on $\la \delta \szz \ra$
then becomes, on large length scales, Poisson's equation:
\be
\nabla^2  \la \delta \szz \ra = \partial_t \la \delta \rho \ra
\ee
which means that the stress propagation becomes somewhat similar to that in an elastic body, where stresses obey 
an elliptic equation of similar type \cite{LL}. In particular, the cone structure of stress
propagation, which is associated with the underlying, hyperbolic, wave equation finally disappears; the average response to a localised perturbation becomes a
broad `bump' of width comparable to the height of the pile. It is thus rather interesting to see that, 
within {\sc mca}, there is a phase transition from a `wavelike' mode to a `diffusive' mode of 
stress propagation; the observation of the `cone' thus requires that the packing is
not too disordered. Certainly for relatively ordered packings the cone exists and has been observed
experimentally \cite{Pouliquen} and numerically \cite{Eloy}. One should however add some remarks:

-- It is possible that the above transition is an artefact, due to the fact that $v$ is taken to be
gaussian, which strictly speaking is not allowed, since the local value of $c_0^2$ should always
be positive. One can show for some other problems of the same type that a similar transition is artificially induced by
the gaussian approximation when it cannot really exist on physical grounds.
In this respect, it is interesting to note that the first non gaussian correction tends to increase $c_R$,
for negative kurtosis as might be expected for a bounded $v$ distribution. 

-- It should be noted that the predicted effective constitutive relation between horizontal and
vertical normal stresses has a negative sign if $c_R^2 <0$. This means that increasing the vertical
stress should reduce the horizontal stress, which is only possible is the grains move. Hence, the region where $c_R^2 <0$ is probably impossible to reach physically: the system will rearrange spontaneously as to reduce the disorder, and to make $c_R^2 \geq 0$. Note however that, as already discussed above, the disorder which results of such a rearrangement might be strongly correlated, and correspond to an arching effect, as in \cite{CB}.

\subsection{The correlation function}

Coming back to the weak disorder case, the major problem for the direct observation of the
`light cone' is the fact that the perturbation representing the point source should be small
(otherwise the packing structure would changes in an inhomogeneous way, thereby affecting the
value of $c_0^2$ in a non uniform way), but large enough for the response to be detected.
A better possibility, as we show now, could be to measure the correlation function of the stress field. 
We again consider the stress correlation function in the case where the mass of each grain is small
($\rho=0$) and a random or a constant overload is applied on the top of the `silo'.
With the new convention for the bar (for $G$) and the cross (for $\partial_t\rho$), the self-consistent
diagrammatic equation (\ref{diagequa7}) is strictly valid in the tensorial case.
When writing it into its usual mathematical form, the only difference with the scalar model
is that now the weight source term is $w(k,0) \de'(t)$, leading to 
$S_0(k,t',t'')=C(k,0) \de'(t') \de'(t'')$. 

The calculation of the correlation function is very similar to the scalar case.
In order to carry out the calculations to the end, we have neglected the dispersion term
$\alpha |k|$ in the expressions for $G$ and $R$. The analogue of equation (\ref{C2}) is now,
for weak disorder,
\bea
\lefteqn { C(k,t) = C(k,0) \cos^2 \left [ c_R kt \right ] e^{-2\ga k^2 t} + } \nonumber \\
\label{C2vect}
& & \sigma^2 {c_0^4 \over c_R^2} k^2 \int_0^t dt'
\sin^2 \left [ c_R k (t-t') \right ] e^{-2\ga k^2 (t-t')} \tilde{C}(t')
\eea
The function
$\tilde{C}(t')=\int {dk \over 2\pi} C(k,t')$ is of identical form to the scalar case; only the
expressions for $a_0$ (for the random overload) and $b_0$ (for the constant overload) are different:
\bea
\label{a0wd}
a_0 & = & {A_0^2 \over 4\sqrt{2\pi\ga}} \cdot
{1 \over 1-{\sigma^2 c_0^4\over 4 \pi \ga^2 c_R^2}
\arctan \left ( {\pi\ga \over c_R} \right ) }\\
\label{b0wd}
b_0 & = & {B_0^2 \over 2\pi} \cdot
{1 \over 1-{\sigma^2 c_0^4\over 4 \pi \ga^2 c_R^2}
\arctan \left ( {\pi\ga \over c_R} \right ) }
\eea
Knowing $\tilde{C}(t')$, $C(x,t)$ can be computed from equation (\ref{C2vect}).
For the case of a constant overload, the shape of the correlation function is very close to
the one showed in figure \ref{CSCfig} for the scalar model.
The case of the random overload however is much more interesting since the fact that
information travels along a cone of angle $c_R$ appears clearly: the correlation function
presents {\it two} peaks. The first one is of course at $x=0$, while the second is at
$x=2 c_R t$, which simply means that the two points at the bottom of the information cone share
the same information coming from the apex of this cone \footnote{In the absence of disorder, the correlation function consists of two $\delta$ peaks, one at $x=0$ and the other at $x=2c_0 t$ of half the amplitude.}.
\bfig[h]
\bc
\epsfysize=8cm
\epsfbox{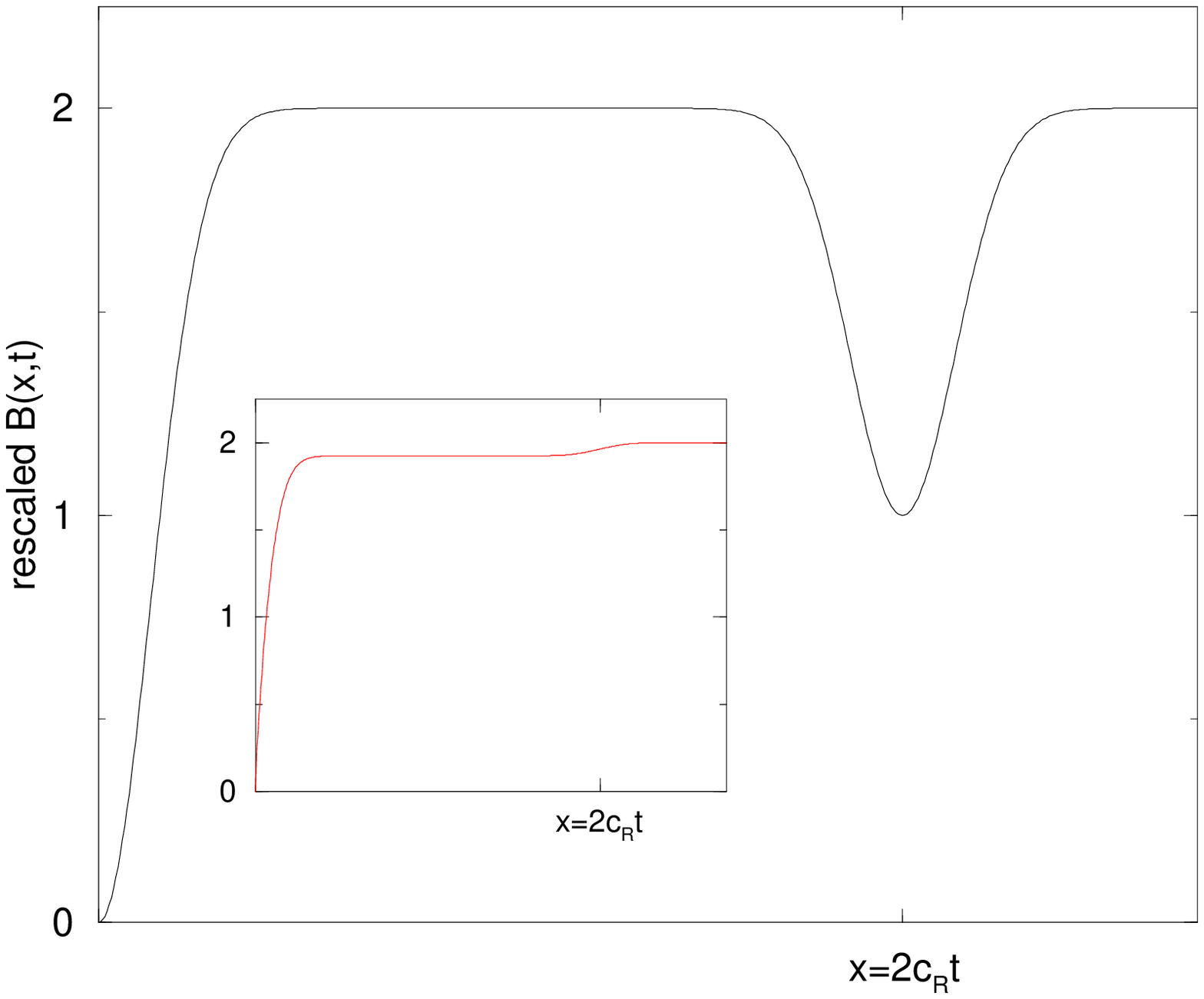}
\caption{\small Correlation function for the case of a random overload. Note the presence of
a peak centered in $x=2 c_R t$ which reflects the fact that information in the tensorial
model is travelling along a cone of angle of $c_R$.
In the case of a fluctuating density
in the bulk of the pile, one should integrate (\protect\ref{Cohv}) with respect to
$t$. The result is plotted in the inset: the correlation reaches rapidly a first plateau, and then increases again to a higher value around $x=2c_R t$. The relative difference of height between the two plateaus decreases as $t^{-1/2}$.
\label{Cvectfig}}
\ec
\efig
If we forget the second term of the
right hand of equation (\ref{C2vect}) which is negligible
compared to the first one at large $t$, we can see that the second peak of the correlation
function has a width $\propto \sqrt{t}$ and a height $\propto 1/\sqrt{t}$.
This approximation is actually equivalent to saying that the (linear) effective equations
(\ref{effequa1}, \ref{effequa2}) are sufficient to calculate the correlation function
for large times. Other source terms, such as a fluctuating density in the bulk of the pile,
can thus be easily accommodated by linear superposition.
We have thus plotted on figure \ref{Cvectfig} the quantity $B(x,t) = C(0,t)-C(x,t)$, omitting the
second term in the r.h.s. of equation (\ref{C2vect}). Analytically, we have
\be
\label{Cohv}
B(x,t)={A_0^2 \over 4 \sqrt{8\pi\ga t}} \left [
2+2e^{-{c_R^2 t \over 2\ga}}-2e^{-{x^2 \over 8\ga t}}-e^{-{(x+2c_R t)^2 \over 8\ga t}}
-e^{-{(x-2c_R t)^2 \over 8\ga t}} \right ]
\ee
This result is of importance since the shape of this correlation function clearly differs from the
corresponding one in the scalar model. Measuring carefully the averaged correlation function of a granular
system could then confirm (or disprove) the presence of a light ray-like propagation. In this respect, it is interesting to plot the correlation function for three dimensional packings as well.
\bfig[ht]
\bc
\epsfysize=8cm
\epsfbox{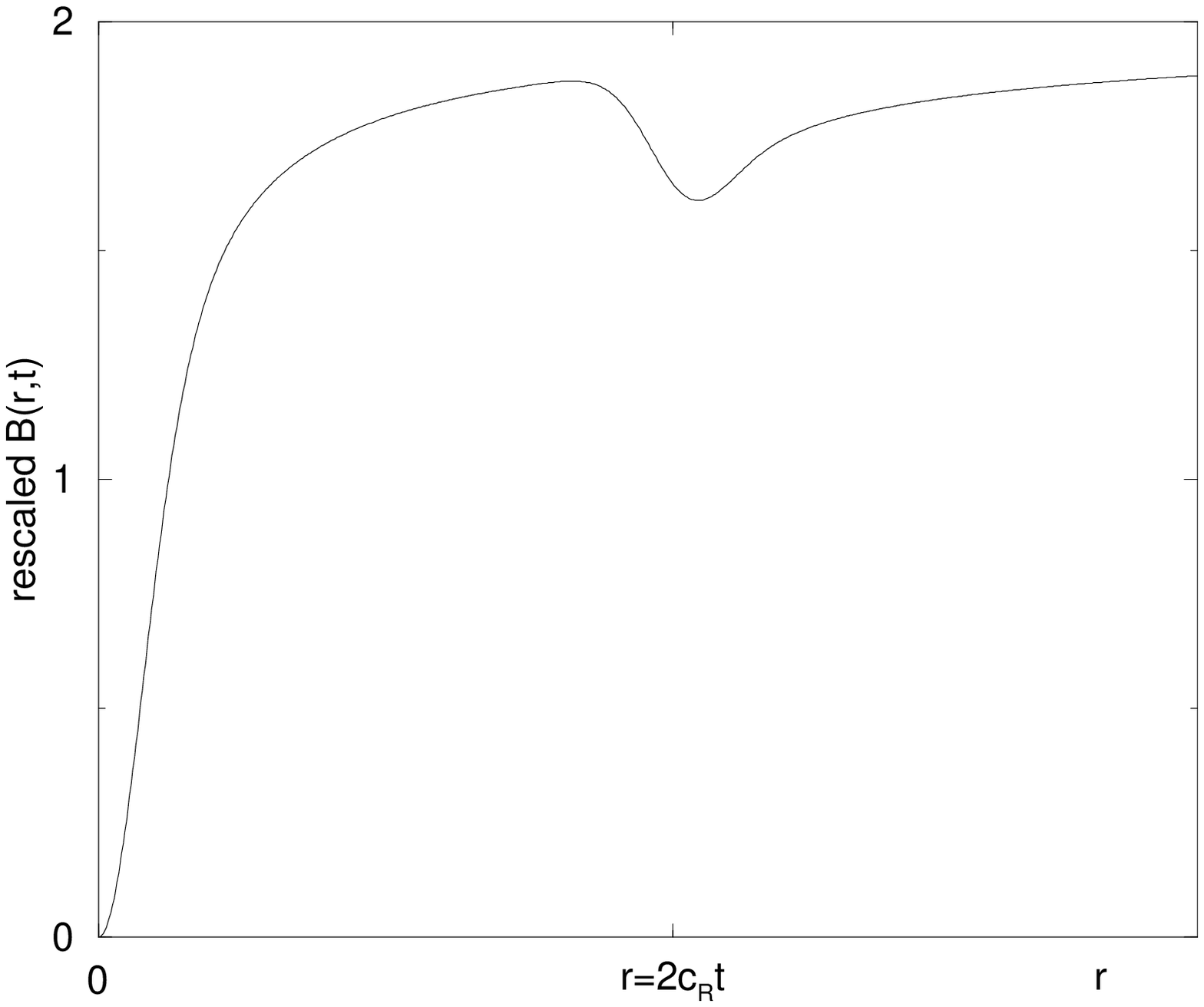}
\caption{\small Correlation function for three dimensional disordered packings with a 
random overload, neglecting again the second term in equation (\protect\ref{C2vect}). Note that, as 
in two dimensions, the correlation function exhibits a peak around $r=2c_Rt$.
\label{corr3D}}
\ec
\efig
This correlation function only depends 
on the radial distance $r$ between the two points, as is plotted in figure \ref{corr3D}. We note that, much as in two dimensions, the correlation decreases sharply on the scale of a few grains, but increases again for distances of the order of the height of the pile. Note
that a stress correlation function was actually recently 
measured in \cite{Nagel} and found to be featureless, but on very short scales $x \leq 5 a$, as compared to the height of the pile $H \simeq 100 a$. We thus expect the features
of the correlation function to show up on much larger scales $\sim 2 c_R H$.

\section{Generalized wave equations}

It is tempting to generalize equations (\ref{equiF1}, \ref{equiF2}) and write the
most general linear equations governing the propagation of the forces
which are compatible with the (local) conservation rules. These equations
were first written by de Gennes \cite{PGDG}:
\bea
\label{equiF1bis}
\partial_t F_t + \partial_x \left[\eta'(x,t) F_x + \mu'(x,t) F_t \right] & = & \rho \\
\label{equiF2bis}
\partial_t F_x + \partial_x \left[\eta(x,t) F_t + \mu(x,t) F_x \right] & = & 0
\eea
Note that the terms $\mu,\,\mu'$ break the symmetry $x \to -x$. This is allowed 
locally and does not show up on large scales if their average 
is zero. Another possibility (but without noise), considered in detail in \cite{FPA},
is that $\mu(x,t)$ changes sign with $x$, i.e.:
$\mu(x,t)=\mu \ \mbox{sign}(x)$, which describes the fact that the texture of a
sandpile depends on which `side' of the pile one is looking at. Interestingly,
equations (\ref{equiF1bis}, \ref{equiF2bis}) still lead to wave-like propagation, but
now the bisector of the `light cone' makes a non zero angle with the vertical
(when $\mu$ or $\mu'$ are non zero). In other words, equations
(\ref{equiF1bis}, \ref{equiF2bis}) describe a situation where not only the opening 
angle of the cone can vary in space, but also its average orientation.

The same analytical techniques as above can be still be used. We shall only
discuss some special cases \footnote{To lowest order in perturbation theory,
the case where disorder in present in the four terms $\eta,\eta',\mu,\mu'$
simultaneously is very simply obtained by adding the corrections induced by each term 
taken individually.}:

$\circ$ Let us first set $\mu=\mu'=0$ and consider the case where $\eta'$ is random,
and $\eta$ fixed (and equal to $c_0^2$). Taking $\eta'(x,t)=\eta'_0 \ (1+v(x,t))$ with
the noise $v$ as above, one finds that the renormalized value of $\eta'$ is:
\be
\eta'_R = \eta_0' \left(1- {c_0^2 \eta_0' \sigma^2 \La^3\ell_t \over 6}\right)
\ee
Now, on large length scales, one must recover the continuum equilibrium equations
for the stress tensor, equations (\ref{equi1}, \ref{equi2}). The condition of zero torque requires
that the stress tensor is symmetric, and thus one must set 
$\eta'_R \equiv 1$, which imposes a relation between $\eta'_0$ and the amplitude
of the noise $\sigma$. Note that beyond a certain value of $\sigma$, this
relation can no longer be satisfied with a real $\eta_0'$. This again means that the
packing is unstable mechanically and will rearrange so as to reduce the disorder.

$\circ$ Another interesting 
class of models, which one can call `$\mu$ models', is such that:
$\eta=c_0^2, \eta'=1$, but $\mu(x,t)=c_0 v(x,t)$ and $\mu'=0$ or vice-versa.
These two cases yield identical results, namely, in the large $t$ limit:
\bea
\label{cas1et4}
R(k,t) & = & \cos \left ( c_0 k t \right ) e^{-\ga k^2 t} \theta(t) \\
\label{cas1et4'}
R_s(k,t) & = & -i c_0 \sin \left ( c_0 k t \right ) e^{-\ga k^2 t} \theta(t)
\eea
where $\ga = {c_0^2 \La \sigma^2 \over 4}$. Note that in these cases, the response peaks
acquire a finite diffusive width $\propto \sqrt{t}$, but the angle of the information cone
is unaffected by the disorder (i.e. $c_0$ is not renormalized).

$\circ$ Finally, there are special `symmetry' conditions
where the equations can be decoupled and reduced to two
`scalar' models. We will refer to this case as the `double scalar' model. This
occurs when $\mu=\mu'=c_0v_1(x,t)$ and $\eta'=\eta/c_0^2=1+v_2(x,t)$ where $v_1, v_2$ are
two different sets of noise. Let us define
$\sigma_\pm = c_0  F_t \pm F_x $, $x_\pm = x \mp c_0 t$ and $v_\pm = v_1 \pm v_2$, we then obtain:
\bea
\label{msc1}
\partial_t \sigma_+ & = & c_0 \rho -c_0 \ \partial_{x_+} [v_+ \sigma_+]\\
\label{msc2}
\partial_t \sigma_- & = & c_0 \rho -c_0 \ \partial_{x_-} [v_- \sigma_-]
\eea
showing that $\sigma_+$ and $\sigma_-$ decouple, each propagating along two rays, of `velocity' $\pm c_0$, 
plus a small noise $v_\pm$ which, as in the scalar case, generates a nonzero diffusion constant. 
The response functions for $\szz$ and $\sxz$ are thus again made of two diffusive peaks of width
$\propto \sqrt{t}$, centered in $x = \pm c_0 t$. The interest of this double scalar limit is
that one can deduce simply the probability distribution of the stresses from the Chicago model.
This is developed below. Note also that by construction, this special form of disorder 
does not lead to negative vertical stresses.

\section{Stress distribution within the tensorial model}
\label{numerics}

A physically relevant question is to know how local stresses are distributed. We have seen above
that within a scalar approach, an exponential-like distribution (possibly of the type $\exp -w^\beta$, with
$\beta \geq 1$) is expected \cite{Liu, Copper}. One can wonder whether this exponential
distribution survives within a tensorial description, and what happens for very small stresses $w \to 0$.
Unfortunately, the full distribution can only be computed analytically for the `double scalar' model;
but numerical results have also been obtained for the random symmetric model, and are described below. 

\subsection{The `double-scalar' limit}

In the `double scalar' limit, the histogram of the stress distribution is obtained trivially
by noting that since $\sigma_+=w_1$ and $\sigma_-=w_2$ travel along different paths, they are independent
random variables. Taking $c_0$ to be unity for simplicity, one thus finds:
\bea
\label{his1}
P(\szz) & = & \int dw_1 \int dw_2 P^*(w_1) P^*(w_2) \delta(\szz-\frac{w_1+w_2}{2}) \\
\label{his2}
P(\sxz) & = & \int dw_1 \int dw_2 P^*(w_1) P^*(w_2) \delta(\sxz-\frac{w_1-w_2}{2})
\eea
where $P^*$ is the distribution of weight pertaining to the scalar case, which, as 
mentionned above, depends on the specific form of the local disorder and on the 
discretisation procedure. In the strong disorder case which leads to equation (\ref{exptail}) [in the case $N=2$], we thus find that $P(\szz)$ is still decaying
exponentially (it is actually a $\Gamma$ distribution of parameter $2N$), although its variance is
reduced by a factor $2$. For $N=2$, one simply gets
\bea
\label{his3}
P(\szz) & = & {8 \over 3} \szz^3 e^{-2\szz} \\
\label{his4}
P(\sxz) & = & \left ( |\sxz| + {1 \over 2} \right ) e^{-2|\sxz|}
\eea
The preexponential factor is therefore noticeable different from the prediction of
equation (\ref{exptail}).

\subsection{Numerical histograms for the random symmetric model and open problems}

The numerical analysis of equations (\ref{equi1}, \ref{equi2}), with a stochastic
constitutive relation $\sxx =\eta [1+v(x,t)]\szz$ is actually not an easy task, and the final results depend rather sensitively on the chosen discretisation. For example, a naive discretisation of the random wave equation leads to a non zero diffusive width even in the absence of disorder, and  thus makes it hard to measure the `true' response function, which should, in the absence of disorder, consists of two $\delta$ peaks. However, it should be noted that such diffusive term (or order $a$) are actually expected physically -- they indeed appear when equations (\ref{wemicro1}, \ref{wemicro2}) are expanded to second order in the lattice spacing. We shall come back to this point below.
\bfig[bht]
\bc
\epsfysize=4cm
\epsfbox{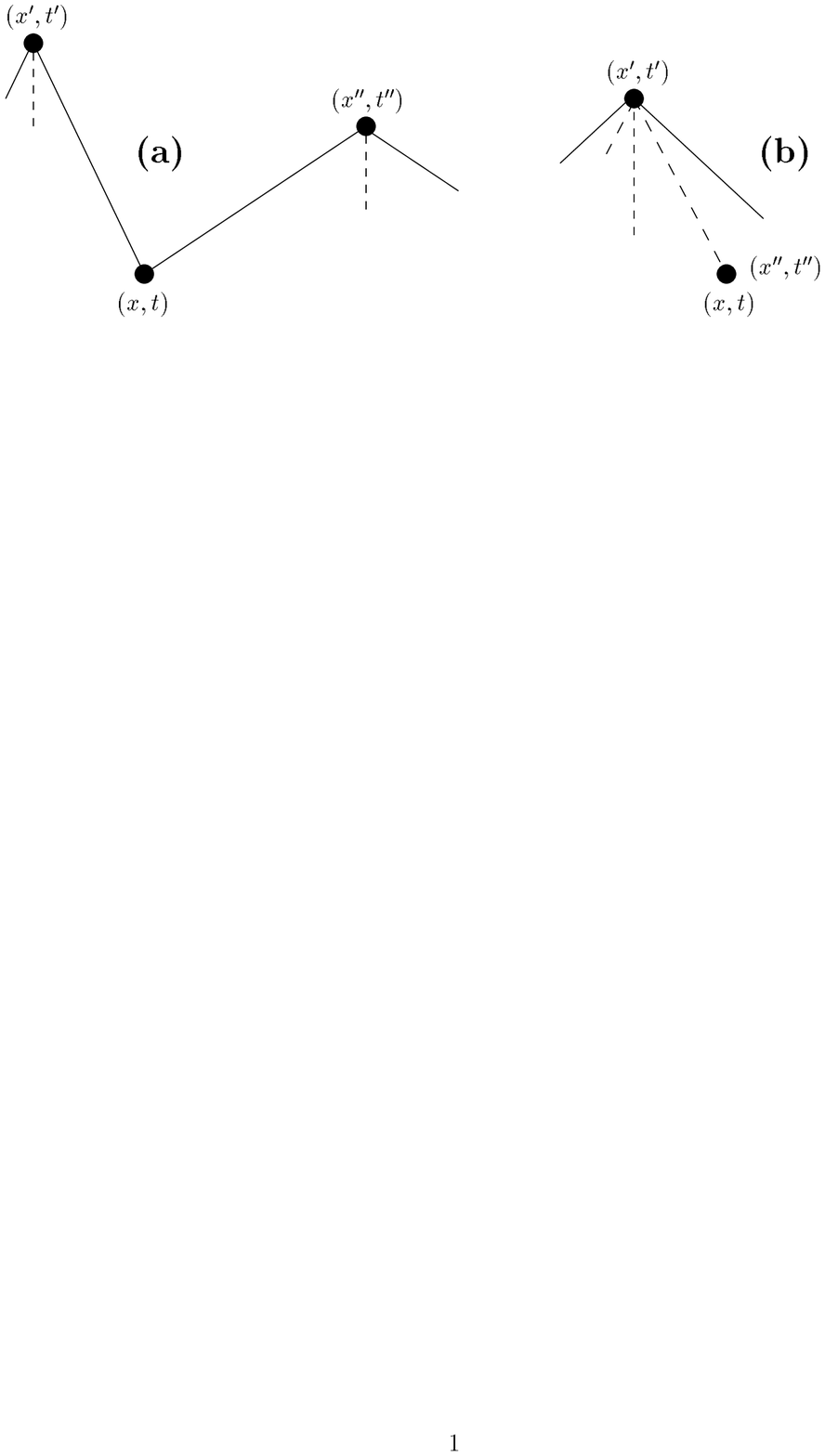}
\caption{\small The left-hand picture (a) shows the construction rule of the characteristics
network: the `child' point $(x,t)$ is located at the intersection of the cones from the two
`parent' points $(x',t')$ and $(x'',t'')$. When the cones do not intersect (b), we choose $(x,t)$
and $(x'',t'')$ to be coincident.
\label{reseau}}
\ec
\efig

The method we chose is the  following. Starting with points regularly arranged at $t=0$,
we construct the network of characteristics (in the mathematical sense).
To each point $(x,t)$ is associated a `speed of light' $c_0(x,t)=c_0 \sqrt{[1+v(x,t)]}$
which determines the directions of lines which propagate the component of the stress parallel to that line, away from the point $(x,t)$.
The point $(x,t)$ is then generated by two `parents' points $(x',t')$ and $(x'',t'')$ as indicated
on figure \ref{reseau}-(a).
It sometimes happens that the cone from $(x',t')$ is so wide that it cannot intersects with
the one from $(x'',t'')$ [see figure \ref{reseau}-(b)]. We then impose $x=x''$ and $t=t''$.
This actually can be viewed as a local kind of arching: the point $(x'',t'')$ not only supports its `parent'
neighbours but also its `same generation' neighbour $(x',t')$.
This method has several advantages. Its physical interpretation is very clear: points 
are
`grains' and characteristics are `stress paths'. Figure \ref{carac} shows an example of the
network of those paths. We can see how stress paths actually merge together and generate arching.
Furthermore, there is strictly no diffusion in the absence of disorder, i.e. the Green function is exactly given by the sum of two $\delta$-peaks.
\bfig[p]
\bc
\epsfysize=6cm
\epsfbox{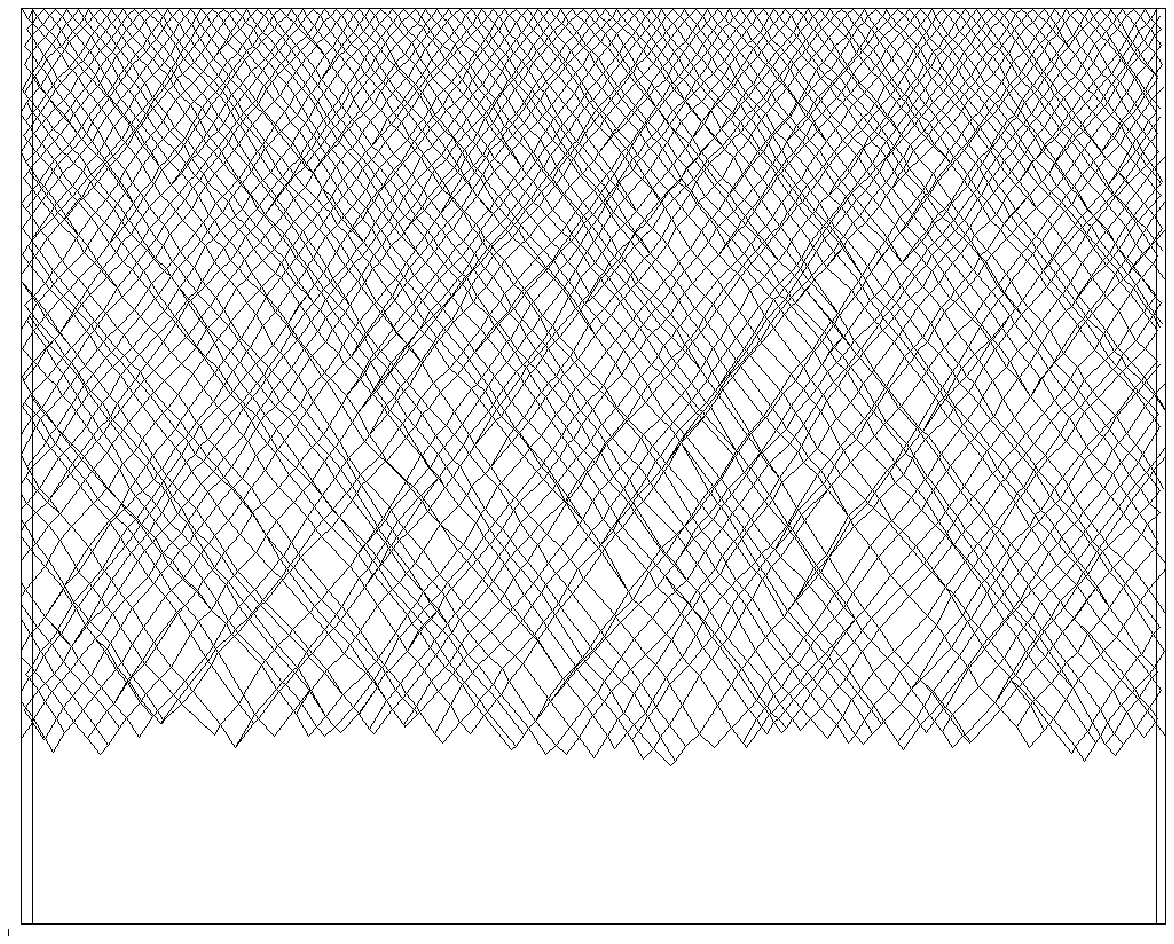}
\caption{\small Stress path network for a periodic silo of width $100\ a$. This picture has been computed
with $\De=0.2$. We have chosen periodic lateral boundary conditions.
\label{carac}}
\ec
\efig
\bfig[p]
\bc
\epsfysize=8cm
\epsfbox{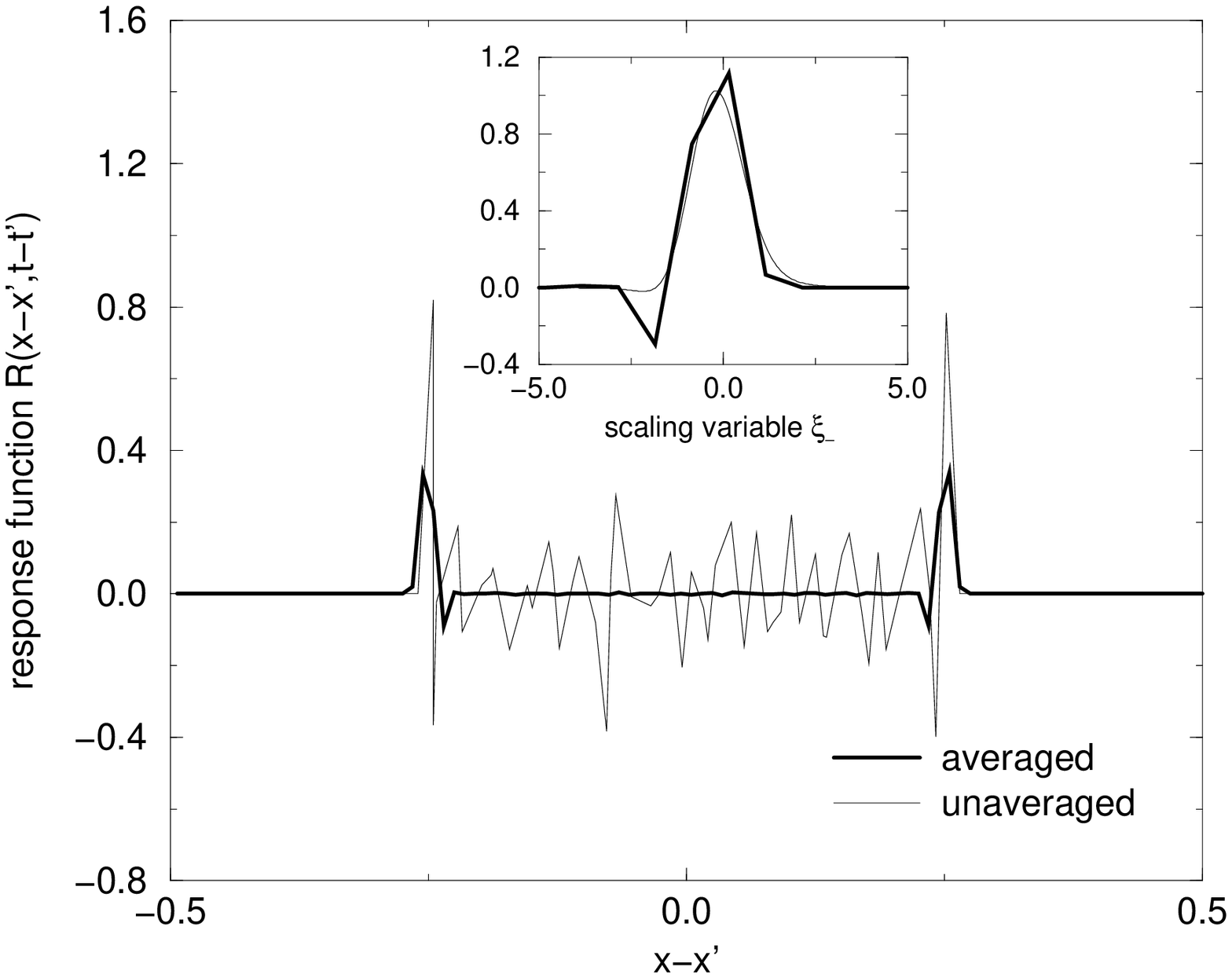}
\caption{\small The main graph shows the response function calculated numerically on a silo of width $100 \ a$
with $\De=0.2$. The thin line is a typical response for a given realisation of the disorder. Note that it takes negative values. The bold line has been averaged
over $5000$ realisation of the disorder. The inset compares the averaged response peak with the one computed analytically, with a negative $\alpha$. Note the negative part, as predicted by the theoretical calculation.
\label{green}}
\ec
\efig

Although the noise $v$ has been implicitly considered to be gaussian throughout this paper,
for numerical simplicity we chose the following algorithm for the calculation of $v(x,t)$.
At each site $(x,t)$, a random angle $\theta$ is uniformly chosen between
$-\De{\pi\over 4}$ and $\De{\pi\over 4}$. $\De$ controls the amplitude of the noise.
We then set $c_0(x,t)=c_0 \tan \left (\pi/4+\theta(x,t) \right)$; 
$v$ and $\theta$ are then related by
$v(x,t)={4\tan\theta(x,t) \over \left (1-\tan\theta(x,t)\right )^2}$.
However, since the lattice itself is
generated by the disorder, the precise correlation function $g_x$ of the $v$'s is not
well controlled in this numerical scheme. This is rather important since
we showed in section \ref{tensorial} that the structure of $g_x$
influences the shape of the response function (it determines whether the negative part of the response lies on the inward or outside edge of the main peak). In fact, the structure of the peaks we obtain numerically
is reversed compared to that of our analytical calculation: see figure \ref{green}. 

The numerical histogram of the force distribution at the bottom of a `silo' 
computed within this numerical scheme immediately reveals some problems.
Since the lattice becomes more distorted as `time' grows, the numerical histogram of
vertical forces keeps broadening and never reaches a stationary shape. Furthermore, 
there is 
a nonzero probability of observing {\it negative} weights which is, as we pointed out already, a structural property of the wave equation with randomness. Clearly, from a physical point of view, this is unacceptable and an additional
rule should be added if the weight becomes locally negative. Some physically
motivated rules could be invented (much as in \cite{CB}), but we do not want to pursue this here, and leave this for future investigations.

In the present paper, we restrict to the case
of a nonzero `bare' diffusion constant which, as argued above, should 
exist on a physical basis. Numerically, we have implemented this is two different ways.
\bfig[bht]
\bc
\epsfysize=8cm
\epsfbox{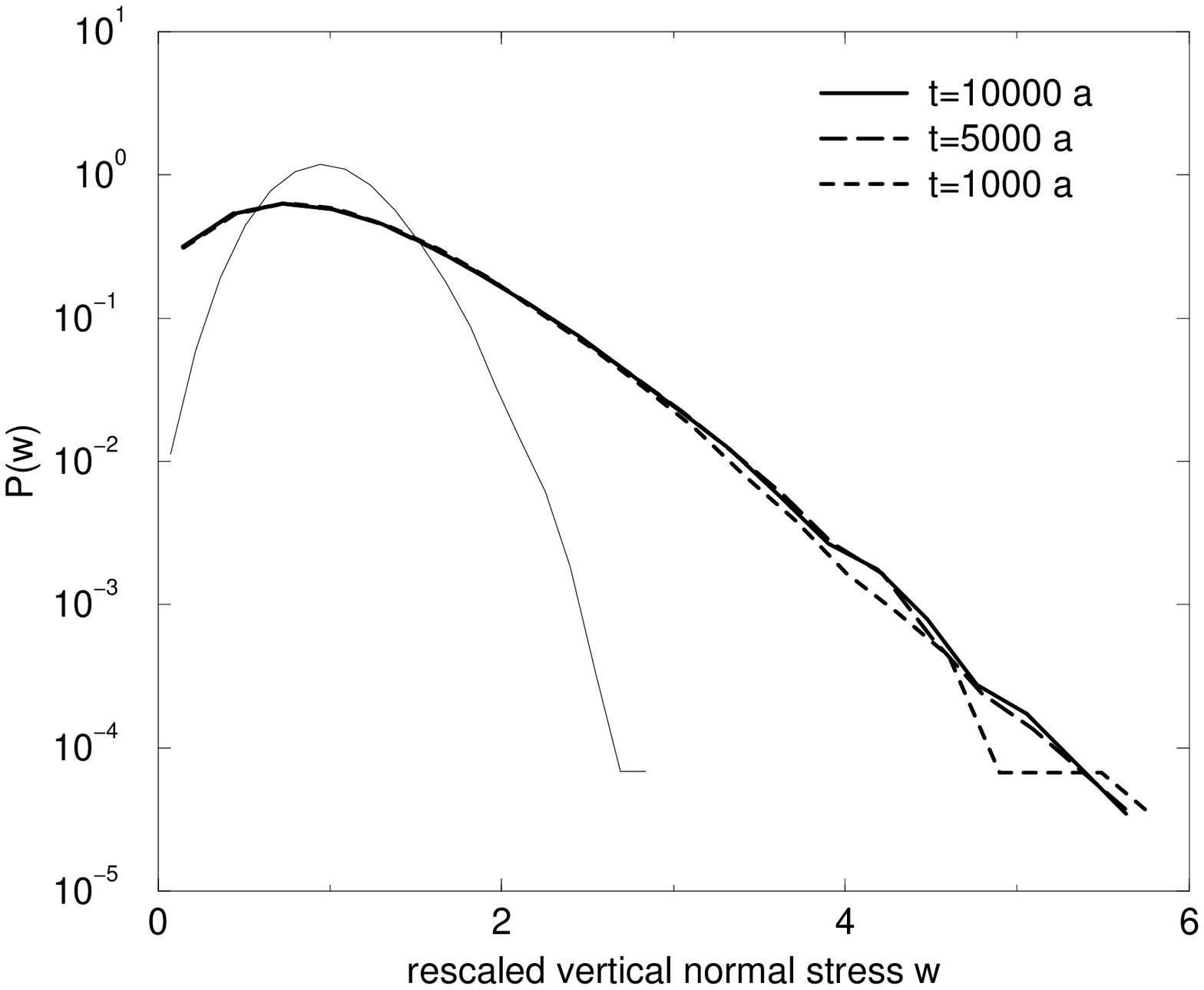}
\caption{\small These curves show the histograms of the vertical normal stress $w$, from which
negative values have been removed. They all have been computed for the three-leg model,
with periodic `silos' of width $1000\ a$.
The three bold (solid, long-dashed and dashed) lines are results from silos where the amplitude
of the noise is maximum ($p_M=1$). The height of those silos is as indicated in the legends.
On the contrary, the thin line represent a $1000\ a \times 1000\ a$ silo where the amplitude of
the noise is $p_M=0.5$ where it is nearly gaussian.
Much like within the scalar model, $P(w)$ shows an exponential tail for large values of
$w$ when the disorder is maximum, while it is better fitted by a stretched exponential,
$\log P(w) \sim -w^\beta$ with $\beta > 1$, for smaller values of $p_M$. 
\label{histo3leg}}
\ec
\efig

$\circ$ The first one corresponds to letting the above 
scheme run until some height $t_D$ and then start afresh with a regular lattice, where the forces are obtained by averaging over the nearest neighbours belonging to the `old' lattice. This averaging procedure is clearly equivalent to
a diffusion term. In this case, the numerical histograms do reach a stationary limit. We note that:

-- The total probability of negative forces is reduced when $t_D$ is smaller.

-- For $t_D \sim a$, the histogram is very nearly gaussian around the average force.

-- For larger $t_D$, the tail of the probability distribution for large 
forces is of the form $\exp -w^\beta$, where $2 > \beta > 1$ (as found in \cite{Eloy}), 
where $\beta$ is decreasing towards one as $t_D$ increases, of for increasing disorder. For $t_D=10\ a$
and $\Delta=0.1$, we found $\beta \simeq 1.6$. The small force region
has a much larger weight than found within the scalar model, although the
presence of negative forces prevents us from being conclusive in this region.

$\circ$ The second scheme consists in simulating directly the three-leg model introduced above, with a
random $p$ chosen between $0$ and $p_M$. These scheme is thus very close in spirit to the Chicago model. Again, the local forces are not everywhere positive, and thus the small force region cannot
be reliably studied. Nevertheless, the large force region, however, behaves much in the same way as in the Chicago model. In particular, as shown in figure \ref{histo3leg}, the tail of the distribution decays as $\exp -w^\beta$, with $\beta \simeq 1$ when $p_M=1$, and with $\beta > 1$ when $p_M < 1$. 

More work is needed to understand the physical implications of the presence
of negative forces and any relation this may have to the static avalanche phenomenon \cite{CB}. However, the above results show that the tail of the force distribution is only exponential in a `strong disorder' limit, where local `arches' (i.e. one grain entirely bearing on a single downward neighbour) has a non zero probability.

\section{Summary-Conclusion}

We have investigated in great detail the r\^ole of a local disorder in the
propagation of stresses in granular media, both within a `scalar' approach,
where only one component of the stress tensor is retained, and within the
full tensorial approach, using a simple linear closure scheme (called `{\sc bcc}' in \cite{WCCB, FPA}),
motivated partly by numerical simulations, which leads to a wave-like equation
for stress propagation. The main effect of this local disorder is, besides introducing
a diffusion like term in the effective, large scale equations, to
renormalise the opening of the angle of the characteristic `light cone' for propagation
of stress. Within a `Mode-Coupling' approximation ({\sc mca}) scheme
(exact for uncorrelated gaussian noise), one finds that this
angle vanishes for a critical disorder, beyond which stress propagates in a
fundamentally different way (this regime might however not be physically relevant). The most 
striking
difference between the scalar and tensorial approach, is the fact
that the response function becomes negative in the latter case, which is a
source of instability of the packing to external perturbations. For moderate
disorder, the response function takes negative values of order one near the point
where the perturbation is applied, and decays with distance. Hence, we expect this instability to occur 
near the point where the perturbation 
is applied; at least near the upper surface of a pile under gravity, the effect occurs for a stress perturbation as small as the weight of one grain, since this is sufficient to 
make the total local vertical stress negative !

Another difference which could be amenable
to experimental verification is the structure of the correlation function,
which gives direct information on how the information travels in the medium.
Because of the analogy between the scalar model and passive scalar convection in 
turbulence, it is furthermore possible that higher
moments of the correlation functions might reveal, in some circumstances, an
intermittent behaviour. Finally, the exponential
fall off of the local stress distribution at high values, first found within the scalar
model, also holds within a tensorial approach, but requires large disorder.

Several open points remain for further studies. First of all, we have only
considered two dimensional packings. The extension to three dimensions is rather 
straightforward -- although the structure of the response functions
becomes inherently more complex in this case (see \cite{BCC}), the main
features discussed here (i.e, diffusive spreading and narrowing of the cone)
are still valid. 

Finally, we have not been able to determine
analytically the histogram for local stresses within the random {\sc bcc} model.
The major unsolved problem is the presence of negative forces, which induce a mechanical
instability, and imposes that an extra rule should be added to the stochastic
wave equation to determine how stress propagates. As emphasized above, we believe that this 
is a direct consequence of the tensorial nature of the problem and can be interpreted as a
signature of ``fragility'' of the contact network, which is generically unstable to very small 
perturbations \cite{CB, rem}.

\vskip 2cm

{\sc acknowledgements:} We thank V. Bucholtz, E. Cl\'ement, J. Duran, C. Eloy and J. Rajchenbach
for discussions.

\end{document}